\documentclass{article}
\usepackage{geometry, booktabs, subcaption}
\usepackage{titlesec}
\usepackage[T1]{fontenc}
\usepackage[utf8]{inputenc}
\usepackage{authblk}
\usepackage{amssymb,amsmath}
\usepackage{bm,bbm,latexsym,mathrsfs,amsmath,amsfonts,amssymb}
\usepackage{graphics}
\providecommand{\keywords}[1]{\textbf{Keywords: } #1}
\titleformat{\section}{\centering\large\scshape}{\thesection}{1em}{}
\titleformat{\subsection}{\centering\normalsize\scshape}{\thesubsection}{1em}{}

\usepackage{hyperref}
\hypersetup{
	colorlinks=true,
	linkcolor=blue,
	filecolor=blue,      
	urlcolor=blue,
	citecolor=blue
}

\usepackage{natbib}
\usepackage{graphicx} 
\graphicspath{{art/}}
\usepackage[table]{xcolor}
\definecolor{lgray}{gray}{0.80}

\usepackage{setspace}
\usepackage{xr}
\usepackage{color}
\usepackage{amsmath}
\usepackage{amsfonts,dsfont,mathrsfs}
\usepackage{booktabs,multirow,array}
\usepackage{bm}
\usepackage[linesnumbered,ruled,vlined]{algorithm2e}

\makeatletter
\renewcommand{\algocf@captiontext}[2]{#1\algocf@typo. \AlCapFnt{}#2} 
\def\@algocf@capt@plain{top}
\renewcommand{\algocf@makecaption}[2]{%
	\addtolength{\hsize}{\algomargin}%
	\sbox\@tempboxa{\algocf@captiontext{#1}{#2}}%
	\ifdim\wd\@tempboxa >\hsize
	\hskip .5\algomargin%
	\parbox[t]{\hsize}{\algocf@captiontext{#1}{#2}}
	\else%
	\global\@minipagefalse%
	\hbox to\hsize{\box\@tempboxa}
	\fi%
	\addtolength{\hsize}{-\algomargin}%
}
\makeatother

\newcommand{\calN}{\mathrm{N}}

\newcommand{\iid}{\stackrel{\mathrm{iid}}{\sim}}
\newcommand{\ind}{\stackrel{\mathrm{ind}}{\sim}}
\newcommand{\uno}{\mathbbm{1}}
\newcommand{\e}{\mathrm{e}}

\newcommand{\Rea}{{\mathbb{R}}}

\newcommand{\diag}{\mathrm{diag}}

\newcommand{\E}{\mathbb{E}}

\newcommand{\virgolette}[1]{``#1''}

\newcommand{\eale}{\color{black}\rm  }

\usepackage{soul}

\newcommand{\D}{\mathrm d}
\newcommand{\bfx}{\mathbf{x}}
\newcommand{\bfy}{\mathbf{y}}

\newtheorem{defin}{\bf Definition}

\renewcommand{\mid}{\,|\,}

\allowdisplaybreaks[3]

\begin{document}

\title{\scshape\LARGE{Clustering blood donors via mixtures of product partition models with covariates}}

\author[1]{Argiento R.\thanks{raffaele.argiento@unibg.it}}
\author[2]{Corradin R.\thanks{riccardo.corradin@unimib.it}}
\author[2]{Guglielmi A.\thanks{alessandra.guglielmi@polimi.it}}
\author[2]{Lanzarone E.\thanks{ettore.lanzarone@unibg.it}}

\affil[1]{\normalsize{Department of Economics, University of Bergamo, Bergamo, Italy and Collegio Carlo Alberto, Torino, Italy}}
\affil[2]{\normalsize{School of Mathematical Sciences, University of Nottingham, Nottingham, UK}}
\affil[3]{\normalsize{Department of Mathematics, Politecnico di Milano}} 
\affil[4]{\normalsize{Department of Management, Information and Production Engineering, University of Bergamo, Bergamo, Italy}} 
\date{\today}

\maketitle 

\begin{abstract}
Motivated by the problem of accurately predicting gap times between successive blood donations, we present here a general class of Bayesian nonparametric models for clustering. These models allow for prediction of new recurrences, accommodating covariate information that describes the personal characteristics of the sample individuals. We introduce a prior for the random partition of the sample individuals which encourages two individuals to be co-clustered if they have similar covariate values. Our prior generalizes PPMx models in the literature, which are defined in terms of cohesion and similarity functions. We assume cohesion functions which yield mixtures of PPMx models, while our similarity functions represent the compactness of a cluster. We show that including covariate information in the prior specification improves the posterior predictive performance and helps interpret the estimated clusters, in terms of covariates in the blood donation application. 

	\vspace{12pt}
	\noindent\keywords{Bayesian cluster models; blood donations; non-exchangeable prior; prediction; random partition; recurrent events.}
\end{abstract}

\section{Introduction}

Blood is an important resource in global health care. To get an idea of its relevance, note that the demand for blood ten years ago was about 10 million units per year in the US and 2.1 million units in Italy \citep{Who2012}, and it is constantly growing. In almost all Western countries blood is collected from donors who give their blood voluntarily and freely. Because it cannot be artificially produced, and its short shelf life limits the time interval between withdrawal and utilization. Therefore, efficient blood stocks and supply chains are of paramount importance for its availability. In modern health care systems, blood is supplied by what is called Blood Donation Supply Chain (BDSC), which accounts for provisioning an adequate amount of blood units to meet the demand of transfusion centers and hospitals. 
Blood collection is the first echelon of the BDSC supply chain, and it  has a relevant impact on the entire system in terms of blood unit flow.   
Some of the problems concerning blood collection from donors, such as demand optimization and scheduling of blood collection centers, have not been thoroughly investigated so far \citep{Bas2018a}.  In particular, a key issue lies in the uncertainty associated with the arrival of donors at the collection centers.
Thus, predicting donations and their temporal distribution is crucial to better feed and control the entire BDSC. 

This work  has been motivated  by applicative  and methodological goals. 
The applicative purpose is the computation of accurate prediction of donation times for the enrolled donors in a  blood collection center. Such prediction is useful for estimating the blood supply of different blood groups and Rh factors over time and for managing the resources. This problem has been faced by predicting the number of donors that will arrive on a given date \citep{bosnes2005predicting}, estimating ARMA models for total  daily blood demand \citep{fortsch2016reducing}, or by analysing blood donor return behaviour using frequentist survival analysis methods \citep{james1996analysis}.    In particular, our data have been collected at the Milano Department of the \emph{Associazione Volontari Italiani del Sangue},  simply referred to as \emph{AVIS} in the following, 
which serves a large hospital located in the same city (Niguarda hospital). The dataset includes donors who did not exit the recurrent donation process in the time window we consider (6.5 years) and, for this reason, we address them as \textit{loyal} donors in the rest of the paper. Moreover, 
AVIS Milano aims   at finding out   if there exist homogeneous clusters of donors, characterized by specific patterns of recurrent donation times and similar characteristics of the donors (given by personal and registry information). Such clustering is relevant for understanding how the donation patterns may vary, given specific characteristics of the donors, but it also  helps to improve the predictions of future donation times.

From the methodological viewpoint, we propose suitable Bayesian models for clustering donors using recurrent event data. The models  allow prediction of  new recurrences, accommodating for covariate information that describes the personal characteristics of the sample individuals. At the same time, we use covariate information of the individuals in the prior distribution of the random partition of the sample individuals themselves.
The Bayesian framework naturally handles model-based clustering assuming that the random parameter of the model includes the partition of the sample subjects \citep{hartigan1990,Quintana_Iglesias03}. We introduce a prior encouraging two subjects to co-cluster a priori if their corresponding covariate values are similar.  To sum up, the methodological goal of this paper is to develop a model for recurrent events data with a flexible non-exchangeable prior for the random partition depending on covariates.

Covariate-dependent priors in a Bayesian nonparametric context are relatively new. The seminal work in this area is \cite{maceachern1999}. However, reference papers with clustering with covariates  are 
\cite{PPMx_JSPI10} and \cite{PPMx_JCGS11}. In these works, the prior on the random partition is given via cohesion and similarity functions. The cohesion function $c$ typically depends only on the cluster size, while the similarity $g$ is a non-negative function that formalizes the similarity among the covariates in the cluster. The covariate-dependent prior is given through a product partition approach as 
\begin{equation}
\label{eq:ppmx}
\Pr(\rho_n = \left\{A_1, \dots, A_{k_n}\right\}) \propto \prod_{j=1}^{k_n} c(A_j)g(\bfx_j^*), 
\end{equation}
where $\rho_n$ denotes the partition of the $n$ sample subjects (cf. Section~\ref{sec:clustering}) and $\bfx_j^*= \{x_i, i \in A_j\}$ denotes the collection of covariates corresponding to items belonging to the $j$-th cluster. 
In \cite{PPMx_JSPI10} and \cite{PPMx_JCGS11}, the cohesion function is derived from 
the Dirichlet process
and  the similarity $g$ is the marginal distribution in an auxiliary probability model, even if the $x_i$, $i =1, \dots, n$, are not assumed random. 
 For similar approaches, possibly including variable selection or 
spatial dependence, see  \cite{park2010bayesian} and \cite{quintana2015cluster}, \cite{barcella2016variable}, \cite{sPPM}, \cite{page2018calibrating}  and \cite{Pag22}.
Alternative models   with dependent priors for random partitions are in \cite{dahl2017random}, \cite{dahl2008distance},  \cite{blei2011distance} and \cite{DPP}. 

Our covariate-dependent prior generalizes \eqref{eq:ppmx} into two directions: $i$)
to mitigate the \textit{rich-get-richer} property,
we depart from the cohesion function of the Dirichlet process, and  assume the cohesion function $c$ generated by a more general class of random probability measures, namely the normalized completely random measures \citep{Reg_etal_2003}; $ii$) we consider similarity functions $g$ which are not marginal densities; borrowing the idea from data-driven clustering approaches, we introduce $g$'s measuring the compactness of covariates in each cluster.   In this paper we use \textit{compactness} to denote a measure of proximity of the covariate vectors in a cluster, i.e., a cluster is compact when the total distance between covariates in the cluster and the associated centroid is \textit{small}.  
The resulting model turns out to be a mixture of PPMx models as in \eqref{eq:ppmx}, allowing the construction of a  general MCMC sampler, which does not depend on the specific choice of similarity.

We first describe the model for a unidimensional regression setting and then consider  a more general model for AVIS data using a longitudinal approach for the sequence of gap times between recurrent events (blood donations). In the latter case, since the gap times (in the log scale) are skewed, we assume a skew-normal distribution for the response \citep{azzalini2005skew, arellano2006unification}.
 We consider three different similarity functions in the simulated examples and the motivating application, discussing how their analytical properties might influence posterior inference. 
Note that, since the analytic normalizing constants of some of the full-conditionals  of our MCMC are unknown, we cannot assume that the hyperparameters 
of the cohesion or of the similarity functions are random.  
However, we discuss how to set these hyperparameters.

The design of an MCMC sampler for the computation of posterior inference is among the contributions of our paper,  also accommodating for
the longitudinal nature of the responses and the skew-normal sampling model for the blood donation application.

We mention here that our prior $\pi(\rho_n)$ has the attractive property of encouraging individuals with equal or  similar covariates to be co-clustered. Nevertheless, this prior  does not have the marginal invariance property, that is, 
the prior of the random partition for $n$ individual cannot be obtained as the marginal of the prior 
of the random partition for $n+1$ individuals.

The paper is structured as follows.  
Section~\ref{sec:clustering} introduces mixtures of product partition models derived by normalized completely random measures.  In Section~\ref{sec:model} 
we propose our covariate-dependent model, while Section~\ref{sec:posterior_calc} reports a few details on posterior calculation.  Section~\ref{sec:similarity_g}  discusses alternative similarity functions. 
We discuss posterior inference for the AVIS data in Section~\ref{sec:avis}. 
Finally, Section~\ref{sec:discussion} concludes the paper with a discussion.

\section{A Bayesian nonparametric framework for  clustering}
\label{sec:clustering}
We consider a Bayesian model in which the latent partition of data is a random variable,
distributed according to a   prior.  Let $\{Y_1,...,Y_n\}$ be a set of observations, where 
$Y_i$ takes values in $\Rea^k$ for some integer $k$, endowed with associated Borel $\sigma$-field; we will not refer explicitly to $\sigma$-fields in this paper, unless this is necessary.  
The typical assumption on its conditional distribution is
\begin{equation}
\label{eq:modello1}
Y_1,...,Y_n\mid A_1,\dots,A_{k_n},\mathbf{\theta}_1^*, \dots, \mathbf{\theta}_{k_n}^*
\sim \prod_{j=1}^{k_n} \left\{\prod_{i\in A_j}^{}f(y_i;\theta^*_j)
\right\},
\end{equation}
where $\rho_n:=\left\{A_1,\dots,A_{k_n} \right\}$ is a partition of 
the data label set $[n] = \{1,\dots,n\}$, i.e., a collection of $k_n$ non-empty disjoint subsets such that their union equals $[n]$, and
$\{f(\cdot;{\theta}), \theta\in\Theta\}$ is a family of densities on the sample space indexed by a parameter  $\theta \in \Theta$, where $\Theta$ is a (measurable) subset of $\Rea^p$ for some integer $p$.  
Observe that here $k_n$ denotes the number of clusters in the partition $\rho_n$. It is apparent from \eqref{eq:modello1} that, conditionally on $\rho_n$, data
are independent between different clusters and are independent and identically
distributed (i.i.d.)
within each cluster.
The model specification is completed by assigning a prior distribution to
$(\rho_n,\bm \theta^*)$. 
In this context, the prior for $\rho_n$ is given by 
\begin{equation*}
\label{eq:eppf}
\pi(\rho_n)=\Pr(\rho_n=\{A_1,\dots,A_{k_n}\})=p(n_1,\dots,n_{k_n}),
\end{equation*}
for some function $p(\cdot)$, where $n_j = \#A_{j}$ denotes the cardinality of the $j$-th cluster in $\rho_n$
and, given $ \rho_n$, the parameters in $\{\theta_1,\dots,\theta_{k_n}\}$ in \eqref{eq:modello1} are  
i.i.d. from some fixed distribution $P_0$ diffuse on $\Theta$.

The random partition $\rho_n$  is exchangeable if its distribution is invariant
under the action of all  permutations of $\{1,\ldots,n \}$. In particular, \cite{pit1995exch} proves that 
the exchangeability holds if and only if $p(\cdot)$ is a symmetric function of its arguments. Furthermore, 
if  $p(1)=1$~and
\begin{equation}
\label{eq:marg_inv}
p(n_1,\dots,n_k)=\sum_{i=1}^{k}p(\dots,n_j+1,\dots)+p(n_1,\dots,n_k,1),
\end{equation}
then  the function $p(\cdot)$ is called 
exchangeable partition probability function (eppf). 
Formula  \eqref{eq:marg_inv} is known as consistency of the eppf or marginal invariance, 
i.e., the probability distribution for partitions of $\{1,\ldots,n\}$ is the same as the distribution obtained by marginalizing out $n + 1$ from the probability
distribution for partitions of $\{1,\ldots,n,n+1\}$.

The practical specification of an exchangeable prior $\pi$ for $\rho_n$ is not a simple task, since $\rho_n$ varies in a finite, but complex, space.  
However, by \cite{Pitman96},  for each diffuse distribution $P_0$ and any eppf $p(\cdot)$, there
exists a discrete random probability measure (r.p.m.) $P$ on $\Theta$, $P \sim\Pi(\cdot;p,P_0)$, 
such that model (\ref{eq:modello1}) under the  specified prior $\pi(\rho_n)$ is
equivalent to
\begin{equation*}
\label{eq:hierarchical}
\begin{split}
& Y_i\mid  \theta_i \ind f(y_i; \theta_i)\ \ \ i=1,\dots,n\\
&  \theta_i\mid P \iid P \ \ \ i=1,\dots,n  \qquad 
P\sim \Pi(\cdot;p,P_0),
\end{split}
\end{equation*}
where $P_0$ represents the expectation of $P$. In this case, the corresponding eppf $p(n_1,\ldots,n_{k_n})$ is given in formula (30) in \cite{Pitman96}. Among the possible choices, an interesting class of discrete random probability measures is given by the class of normalized completely random measures \citep{Reg_etal_2003}. 
This family of processes can be defined as 
\begin{equation}
P    
=\sum_{i=1}^{+\infty}\frac{J_i}{T}\delta_{\tau_i}
\label{eq:Norm_Def}
\end{equation}
where 
$\{J_i\}_{i\geq 1}$ are the points of a
Poisson process on $\Rea^+$ with 
intensity $\kappa\rho(s) ds$, with
\begin{equation*}
\int_{0}^{+\infty} \min\{1,s\} \rho(s)ds<+ \infty, \quad \int_{0}^{+\infty}  \rho(s)ds=+\infty,
\label{eq:regolarita}
\end{equation*}
and $\kappa>0$
and $T:=\sum_{i\geq 1} J_i$. The random variables $\tau_i's$ in \eqref{eq:Norm_Def} are i.i.d. from $P_0$ and independent of the 
$J_i$'s. 
The corresponding eppf assumes the following form: 
\begin{equation}
\pi(\rho_n)= 
\int_0^{+\infty}D(u,n) \prod_{j=1}^{k_n}c(u, n_j)  du
\label{eq:Norm_eppf}
\end{equation}
where 
\begin{equation}\label{eq:Du_cu}
D(u,n) = \frac{u^{n-1}}{\Gamma(n)} \exp\left\{-\kappa \int_0^{+\infty} \left( 1-\e^{-us}\right) \rho(s) d s\right\}, \quad c(u,n_j)=  \int_0^{+\infty}\kappa s^{n_j} \e^{-us}\rho(s)  ds,
\end{equation}
See \citet{Pit03} for more details on the expressions in~\eqref{eq:Du_cu}. 


In the next section, we define a new prior for the random partition $\rho_n$ generalizing \eqref{eq:Norm_eppf}. 
We relax the exchangeability condition incorporating covariates in  the prior specification,  so that 
subjects with equal or similar covariates are 
a priori more likely to co-cluster than others. In this way, we extend  product partition models with covariates \citep[PPMx,][]{PPMx_JCGS11}.

\section{Bayesian covariate driven clustering}
\label{sec:model}

In a regression context, let $y_i$ be the response variable and let $\bfx_i\in\Rea^m$  be the covariate vector of the $i$-th observation.  
However, our modelling setup works also in   the more general case when the responses is multidimensional, as we assume in Section~\ref{sec:avis}. 
We denote by $\bfy_j^*$ (or $\bfx_j^*$) the set of all responses $y_i$ (or covariates $\bfx_i$) in cluster $A_j$, with $\bfy^*_j = \{y_i, i \in A_j \}$ (equivalently $\bfx^*_j = \{\bfx_i, i \in A_j \}$). 
As in \eqref{eq:modello1}, we assume that data are independent across groups, conditionally on covariates and the cluster specific parameters, and they are distributed according a regression sampling model 
$f(\cdot;{\mathbf x},\theta)$. The cluster specific parameters are assumed i.i.d from a base distribution $P_0$.
The prior on the partition depends on covariates through a \textit{similarity function}.  
We then assume
\begin{align}
\label{eq:lik1}
Y_1, \dots Y_n \mid  \mathbf{x}_1, \dots,\mathbf{x}_n, \mathbf{\theta}_1^*, \dots, \mathbf{\theta}_{k_n}^*, \rho_n &\sim \prod_{j=1}^{k_n} f(\mathbf Y_j^*\mid \bfx^*_j, \mathbf{\theta}_j^*)\\
\label{eq:distinct}
\mathbf{\theta}_1^*, \dots, \mathbf{\theta}_{k_n}^* \mid  \rho_n &\iid P_0\\
\label{eq:rho_x}
\Pr(\rho_n=\{ A_1,\ldots,A_{k_n}\}\mid \mathbf{x}_1, \dots,\mathbf{x}_n) &\propto 
\int_0^{+\infty}D(u,n) \prod_{j=1}^{k_n}c(u, n_j)g(\bfx^*_j)du,
\end{align}
where $f(\mathbf Y_j^*\mid \bfx^*_j, \mathbf{\theta}_j^*)=\prod_{i\in A_j}^{}f(y_i;{\mathbf x}_i,\theta^*_j)$, 
$n_j$ is the size of cluster $A_j$,  $g(\bfx_j)$ is the similarity function on cluster $A_j$, and $P_0$ is a probability measure diffuse on $\Theta$. Here $D(u,n)$ and $c(u, n_j)$ are those defined in~\eqref{eq:Du_cu}.

The likelihood specification in \eqref{eq:lik1} may be
any model, from simple regression models as in 
Section~\ref{sec:ppmx_gibbs} of the Appendix to the more complex models for gap times of recurrent events as the case study of Section~\ref{sec:avis}. 
The prior \eqref{eq:rho_x} is a perturbation of prior \eqref{eq:Norm_eppf}
in Section~\ref{sec:clustering}: when $g\equiv 1$, i.e., there are no extra information from covariates, the prior mass of each cluster depends only on its size through $c(u,n_j)$, and the prior in \eqref{eq:rho_x} coincides with \eqref{eq:Norm_eppf}; when $g$ is a proper function of $\bfx_j$,
the prior mass associated to the $j$-th cluster increases with $g(\bfx_j)$. 
We remark that a distribution as in~\eqref{eq:rho_x} is well defined, up to a normalization constant that depends on the covariates and the function $g$. By assuming that $g$ takes values in $(0,1]$, we have that 
\[
\begin{split}
M_g(\bfx_1, \dots, \bfx_n) = \sum_{\rho_n}\int_0^\infty D(u, n) \prod_{j=1}^{k_n}c(u, n_j)g(\bfx^*_j)du 
\leq \sum_{\rho_n}\int_0^\infty D(u, n) \prod_{j=1}^{k_n}c(u, n_j)du = 1.
\end{split}
\]
and hence $\pi(\rho_n \mid \bfx_1, \dots, \bfx_n)$ in \eqref{eq:rho_x} 
is well defined.

As a keypoint, we observe that  equation \eqref{eq:rho_x} is a mixture, with respect to $u$, of a product partition model with covariates. In particular, conditionally to $u$, the cohesion function is given by $c(u,n_j)$ and the marginal density of the mixing variable is 
\begin{equation*}
p(u\mid\bfx_1, \dots, \bfx_n) = \frac{D(u,n)}{M_g(\bfx_1, \dots, \bfx_n)} \sum_{\rho_n} \prod_{j=1}^{k_n} c(u, n_j) g(\bfx_j^*).
\label{eq:mar_u}
\end{equation*}
These comments  justify the name  PPMx-mixt for the prior  \eqref{eq:rho_x}. 
Note that covariates that enter in the similarity function do not need to be necessarily the same as those in the regression part of the likelihood, but they can be selected specifically for each application.

For the sake of concreteness, the presentation focuses on a cohesion functions  arising from a specific normalized completeley random measure, 
the normalized generalized gamma process, 
denoted by $\mathrm{NGG}(\kappa, \sigma, P_0)$. Such a choice recovers as particular cases several models commonly used in the Bayesian nonparametric literature, such as the Dirichlet process \citep{Ferguson73}, the normalized inverse-Gaussian process \citep{Lijoi_etal05} and the normalized $\sigma$-stable process  \citep{Pit03}. When a NGG process is used directly as a mixing measure in a mixture, it induces a prior on the number of groups
which is more disperse than the one from the Dirichlet process, allowing us for a more flexible model specification. 
The intensity of the NGG process is given by  
${\displaystyle \rho(ds) = \frac{1}{\Gamma(1-\sigma)} s^{-1-\sigma}\e^{-s}\uno_{(0,+\infty)}(s) ds}$, 
where $\sigma\in [0,1)$ is a discount parameter. In this case
the cohesion function equals
\begin{equation}
c(u,n_j) = \frac{\kappa \ \Gamma(n_j-\sigma)}{\Gamma(1-\sigma)}\frac{1}{(1+u)^{n_j-\sigma}}.
\label{eq:NGG_cohesion}
\end{equation}
Parameter $\sigma$ has a strong impact on the clustering.
In particular, the larger it is, the more disperse is the distribution on the number of clusters. 
This feature mitigates the annoying \textit{rich-get-richer} effect, 
typical of the Dirichlet process, leading to more size-balanced clusters. 
For more details on the behavior of $\sigma$ in NGG's, see for instance 
\cite{Lijoietal07}, \cite{Arg_etal10} and \cite{peps}. 


\section{Posterior calculations}
\label{sec:posterior_calc}
In this section we sketch the Gibbs sampler P\'olya urn scheme for model \eqref{eq:lik1}-\eqref{eq:rho_x}. 
The joint law of data and all parameters, included the auxiliary variable $u$, is given by

\begin{equation}
\label{eq:joint}
\begin{split}
\mathcal{L}\left(\{y_i\}_{i=1}^n, \rho_n,\bm{\theta}^*, u \mid  \left\{\mathbf{x}_i\right\}_{i=1}^n \right) = 
\frac{D(u,n)}{M_g(\bfx_1, \dots, \bfx_n)} \prod_{j=1}^{k_n}  f(\mathbf{y}_j^*\mid \mathbf{x}_j^*, \mathbf{\theta}_j^*)P_0(\mathbf{\theta}_j^*) c(u, n_j) g(\bfx_j^*).
\end{split}
\end{equation} 
 The corresponding algorithm extends the augmented marginal Gibbs sampler for normalized completely random measures mixture models \citep{LijPru_beyond,FavTeh13}.   
We repeatedly sample the full-conditionals below; see Section~\ref{sec:ppmx_gibbs} of the Appendix for full details.
\begin{itemize}
\item The full-conditional  of the mixing parameter $u$, given $\rho_n$, does not include terms which depend on covariates. In particular, we have
\begin{equation*}
\mathcal{L}(u\mid-)= p(u\mid\rho_n,  \mathbf{x}_1, \dots,\mathbf{x}_n) \propto \frac{u^{n-1}}{(1 + u)^{n - \sigma k_n}} \e^{-\left(\kappa \frac{(u+1)^{\sigma}-1}{\sigma} \right)}, \quad u>0.
\end{equation*}

\item For each $j=1,\ldots,k_n$, we independently sample from 
\begin{equation}
\label{eq:fullcond_theta}
{\mathcal L}(\theta_j^*| -)\propto f({\mathbf{y}}_j^*\mid {\mathbf{x}}_j^*, {\mathbf{\theta}}_j^*)P_0(d\mathbf{\theta}_j^*)= \prod_{i\in A_j}f(y_j^*\mid {\mathbf{x}}_j^*, {\mathbf{\theta}}_j^*)P_0(d\mathbf{\theta}_j^*)
\end{equation}
If $f$ and $P_0$ are conjugate, this step is straightforward. If not, we resort to a different sampling strategy such as, e.g., the algorithm in Section~\ref{sec:avis_gs} of the Appendix. 

\item  We update the latent partition the random partition $\rho_n$ using a Gibbs sampling step where the cluster assignment of each item $i$ is updated once at a time.  
We denote by $\rho_{n-1}^{(-i)}$ the partition of $n-1$ items where the $i$-th item has been removed 
and by $\{ i\in A_j\}$ the event that the $i$-th item is assigned to cluster $j$, where $j$ varies 
	in $\left\{1, \dots,k_{n-1}^{(-i)},k_{n-1}^{(-i)}+1\right\}$; $k_{n-1}^{(-i)}$ is the number of clusters available in the partition without $i$.
	Note that $k_{n-1}^{(-i)}+1$ is included to consider the case where the item forms a new cluster. We have
	\begin{equation}\label{eq:cluster_assign}
	 \Pr\left(\{ i\in A_j\}\mid u, \{\mathbf x_i\}_{i=1}^n, \left\{y_i\right\}_{i=1}^n, \rho_{n-1}^{(-i)} \right) 
	\propto \frac{  m(\bfy^*_j \cup\{ y_i\})}{m(\bfy^*_j) }   \frac{c(u,n_j+1) g(\bfx^*_j \cup \{ x_i\}) }{ c(u,n_j) g(\bfx^*_j )},
	\end{equation}
	where $m(y)$ is the marginal density of the parametric Bayesian model defined in \eqref{eq:fullcond_theta}. This density is available analytically in case $f$ and $P_0$ are conjugate. If not, we need to modify this step; see Section~\ref{sec:avis_gs} of the Appendix. 
	In the case of $j = k_{n-1}^{(-i)}+1$, \eqref{eq:cluster_assign} gives the probability of a new cluster as proportional to $m(y_i) c(u,1)$. 
\end{itemize}

The lack of marginal invariance of the prior for the random partition prevent us to compute posterior predictive distributions for new individuals as the integral of the sampling model with respect to the posterior distribution. However, we deal with this calculation considering the responses of new individuals as missing data and including the associated new covariates in the set of all covariates values.
For an alternative approach, based on an importance sampling re-weighting step, see \cite{PPMx_JCGS11}.

%

\section{The choice of the similarity function}
\label{sec:similarity_g}

We remind that in this paper we mean that a cluster is \textit{compact} when the total distance between covariates in the cluster and the associated centroid is \textit{small}.  
We consider similarity functions $g$, with $0<g\leq 1$ (see Section~\ref{sec:model}), that quantifies  the compactness  of the cluster through covariates.    
To this end, let us denote $g(\bfx^*_j):= g(\mathcal{D}_{A_j})$, where   
\begin{equation}
\label{eq:d_cluster}
\mathcal{D}_{A_j} =  \sum_{i \in A_j}d(\mathbf{x}_i, \mathbf{c}_{A_j}),
\end{equation} 
$d$ is a distance between vectors and  
$\mathbf{c}_{A_j}$ is the centroid of the set of covariates in cluster $j$, here assumed as   the Fr\'echet mean. We assume that $g$ is a decreasing (i.e., non-increasing) function of $\mathcal{D}_{A_j}$, so that the smaller is $\mathcal{D}_{A_j}$ (and hence the more compact is the cluster $A_j$), the larger is the value 
$g(\bfx^*_j)$. 
We let ${\mathbf{x}}_i=({\mathbf{x}}^c_i, {\mathbf{x}}^b_i)$, where ${\mathbf{x}}^c_i$ are the available continuous covariates and  ${\mathbf{x}}^b_i$ the available binary covariates. We define the function $d(\cdot, \cdot)$ in \eqref{eq:d_cluster} as 
\begin{equation}
d({\mathbf x}_1,{\mathbf x}_2)=\frac{m_c}{m}d_c({\mathbf{x}}^c_1,{\mathbf{x}}^c_2)+ \frac{m_b}{m} d_b({\mathbf{x}}^b_1, {\mathbf{x}}^b_2),
\label{eq:sum_of_dist}
\end{equation}
where $m_c$, $m_b$ and $m$ denote the number of continuous covariates, the number of binary covariates and the number of total covariates respectively, while $d_c$ and $d_b$ denote the Mahalanobis distance and    $d_b$ is the normalized Hamming distance. 

We propose a list of similarity functions that has proved to work reasonably well in practice: ($i$) ${\displaystyle g_A(\bfx^*_j; \lambda) = \e^{-t^\alpha}}$, for $\alpha>0$; ($ii$) ${\displaystyle g_B(\bfx^*_j; \lambda) = \e^{-\alpha \log(1 + t)}}$, for $\alpha>0$; ($iii$) $g_C(\bfx^*_j; \lambda)=\e^{-t \log(1 + t)}$. Here $t = \lambda \mathcal{D}_{A_j}$. 

\begin{figure}
	\centering
	\includegraphics[width = 0.49\textwidth]{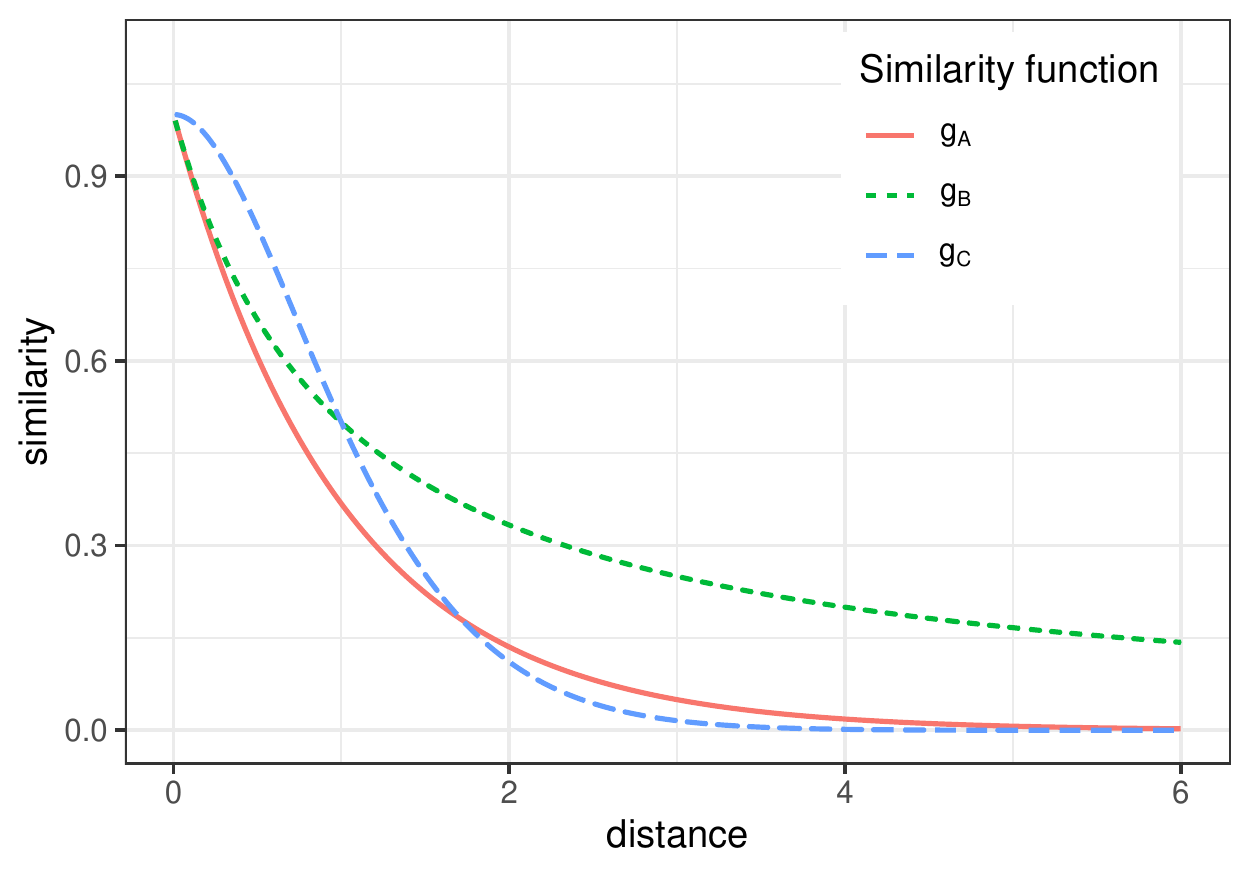}
	\includegraphics[width = 0.49\textwidth]{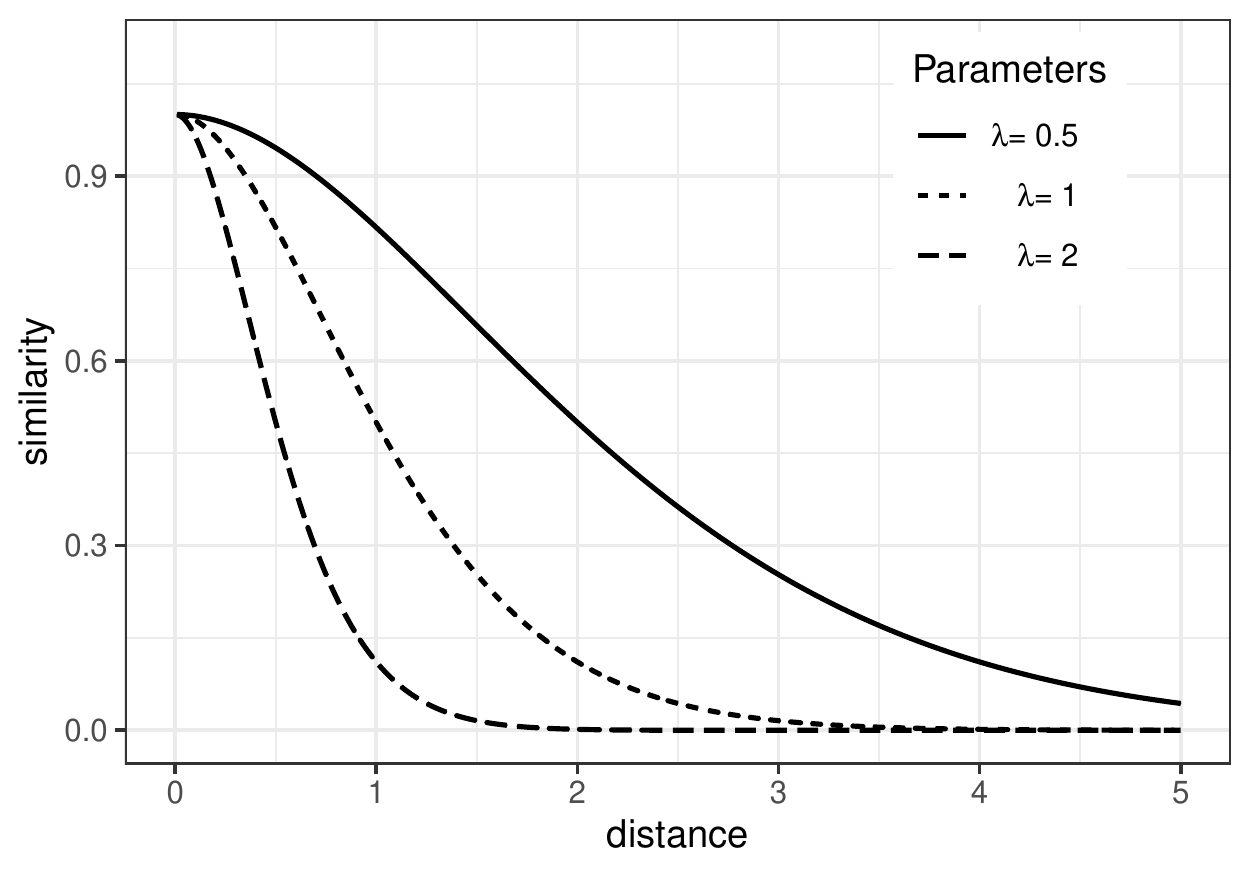}
	\caption{Left: plot of the similarity functions for $\alpha=1$ and $\lambda=1$; right: plot of $g_C$  for different values of $\lambda$.}
	\label{fig:similarities}
\end{figure}

Hyperparameter $\lambda$ is responsible for rescaling the range of values of $\mathcal{D}_{A_j}$ where we evaluate the similarity function. It is the analogue to the \textit{temperature} parameter defined in \cite{dahl2017random} and it tempers how covariates impact on the prior.
Similarly, the power parameter $\alpha$ drives the influence of the covariates in the prior of the random partition, by stretching or compressing the function over its support. Typical vaules for $\alpha$ are $1/2,1,2$. Figure~\ref{fig:similarities} shows the graphs of the three similarities as a function of $t\geq0$. Similarity functions $g_A$ and $g_B$ are intuitive, i.e., their behaviour for $t\rightarrow +\infty$ is exponential and polynomial, respectively. As far as $g_C$ is concerned, we have proposed the expression $\e^{-t\log (1 + t)}$ in such a way that, for large $t$, we contrast the asymptotic 
behavior of the Gamma function in the cohesion \eqref{eq:NGG_cohesion} induced by the NGG. 
Note that, $\mathcal{D}_{A_j\cup \{i\}}\geq \mathcal{D}_{A_j}$ where $\{i
\}$ is a singleton; see Section~\ref{app:secA} of the Appendix, equation~\eqref{eq:inc_property}. 
This imply that the function $g$ penalizes large clusters that are not compact at the same time. This is exactly the feature we would like to guarantee in order to mitigate the rich-get-richer property of the cohesion function associated to the Dirichlet process.  
 
We propose an heuristic strategy to fix $\lambda$: given the available data, we estimate the increment of $\mathcal{D}_{A_j}$ when we add the new observation $\{ x_i\}$ across all possible values of the sample size of $A_j$. For instance, for any sample size $n_j$ from 2 to $n$, we uniformly choose a cluster $A_j$ of size $n_j$, and we add a point $i$ (not in $A_j$),  to obtain a Monte Carlo estimate of the increment $(\mathcal{D}_{A_j  \cup \{i\}} - \mathcal{D}_{A_j})$. We average over the sample size $n_j$, obtaining and estimate $\hat\varepsilon$. Then, we choose $\lambda$ such that $\lambda \hat\varepsilon = \varepsilon^*$, for small values of $\varepsilon^*$, i.e., $\varepsilon^*= 10^{-1},  10^{-2},  5 \times 10^{-3}$.  
Note that the choice of $\varepsilon^*$ and consequently of $\lambda$ calibrates the influence of the similarity function in the posterior estimated clusters, which might be over-driven by the covariates values. For a thorough discussion about calibration of similarities in PPMx models, see \cite{page2018calibrating}.

The specification of a PPMx-mixt prior consists in choosing a cohesion function, $c(u;n_j)\geq 0$ for $n_j \in \{1, \ldots , n\}$, and a nonnegative similarity function $g$. The former quantifies the probability mass of a generic cluster $A_j$, $j = 1,\dots,k_n$, through its cardinality, conditioning to $u>0$ and regardless the knowledge of the subjects in $A_j$, while the latter formalizes the similarity of the covariates.  
The similarity function affects posterior inference through the predictive \eqref{eq:cluster_assign}. The formula shows that there two factors, the first depending only on the responses $y_i$'s through the marginal density of the parametric Bayesian model defined in \eqref{eq:fullcond_theta}, while the second factor depends on the cohesion and the similarity, i.e., 
\begin{equation*}
\frac{c(u,n_j+1) g(\bfx^*_j \cup \{ x_i\}) }{ c(u,n_j) g(\bfx^*_j )}.
\end{equation*}

Observe that the ratio between cohesions $c(u,n_j+1)$ and $c(u,n_j)$ assume values proportional to $n_j-\sigma$, and consists in the usual predictive weight in NGG mixture models. Henceforth, we focus on the above ratio between the similarity values. For $\lambda$ fixed as we have described above, let $t=\lambda\mathcal{D}_{A_j}$ and 
 $t+\varepsilon = \lambda\mathcal{D}_{A_j  \cup \{ i\}} $, so that  
$\varepsilon$ represents the increment of the average center-based distance when $\{ x_i\}$ is assigned to cluster $A_j$. So, it is interesting to study  $g(t+\varepsilon)/g(t)$, for $t>0$ and any fixed $\varepsilon>0$.
This ratio is smaller or equal  than 1, since $g$ is non-increasing. It is advantageous to have a ratio that assumes small values when $t$ is large,  to discourage non-compact clusters. The function $g_C$ is the only one, among the three similarities proposed here, to fulfill this requirements, as shown in Figure ~1 in  Section~\ref{app:secB} of the Appendix. From the same figure, it is  apparent that the ratio is constant  for  $g_A$ and it is increasing  for $g_B$. 
Similarly, it is interesting to study the ratio $g(t+\varepsilon)/g(t)$ also as function of $\varepsilon > 0$, for any fixed value of $t > 0$ which, of course, is non-increasing with $\varepsilon$. 
Hence, when we add observation $i$ in cluster $A_j$, two scenarios can occurr: 
\begin{enumerate}
\item the new observation is similar to the others belonging to $A_j$, so $\varepsilon$ is small, and the ratio is close to one, yielding to a weak penalization of  the weight of the cluster $A_j \cup \{i\}$; 
\item  the new observation strongly differs from the elements in $A_j$, so $\varepsilon$ is large and the ratio becomes small. In this case the model strongly penalizes the weight of the cluster $A_j \cup \{i\}$. 
\end{enumerate}
 
A simulation study to compare the effect of the three similarity functions on posterior  distribution 
is shown in Section~\ref{sec:sim1} of the Appendix. Moreover, a comparison with alternative models using benchmark data is given in  Section~\ref{sec:sim2} of the Appendix

\section{Blood donation data application}
\label{sec:avis}


Our data concern new donors of whole blood donating between January 1st, 2010 and May 15th, 2016 in the main building of AVIS Milano. By a new donor we mean a blood donor who donated for the first time after January 1st, 2010.  Data are recurrent donation times, with extra information summarized in a set of covariates, collected by AVIS physicians. Donors include only \textit{loyal} individuals, i.e., a new donation is expected within a finite amount of time with probability one. 
The resulting dataset contains $11\,505$ donations, made by $2\,912$ donors; the number of recurrent donations for each donor, including the first one, is between $2$ and $21$, so that the number of gap times between recurrent donations vary from $1$ to $20$.

\begin{figure}[!htb]
	\centering
	\includegraphics[width=0.39\textwidth,height=0.3\textwidth]{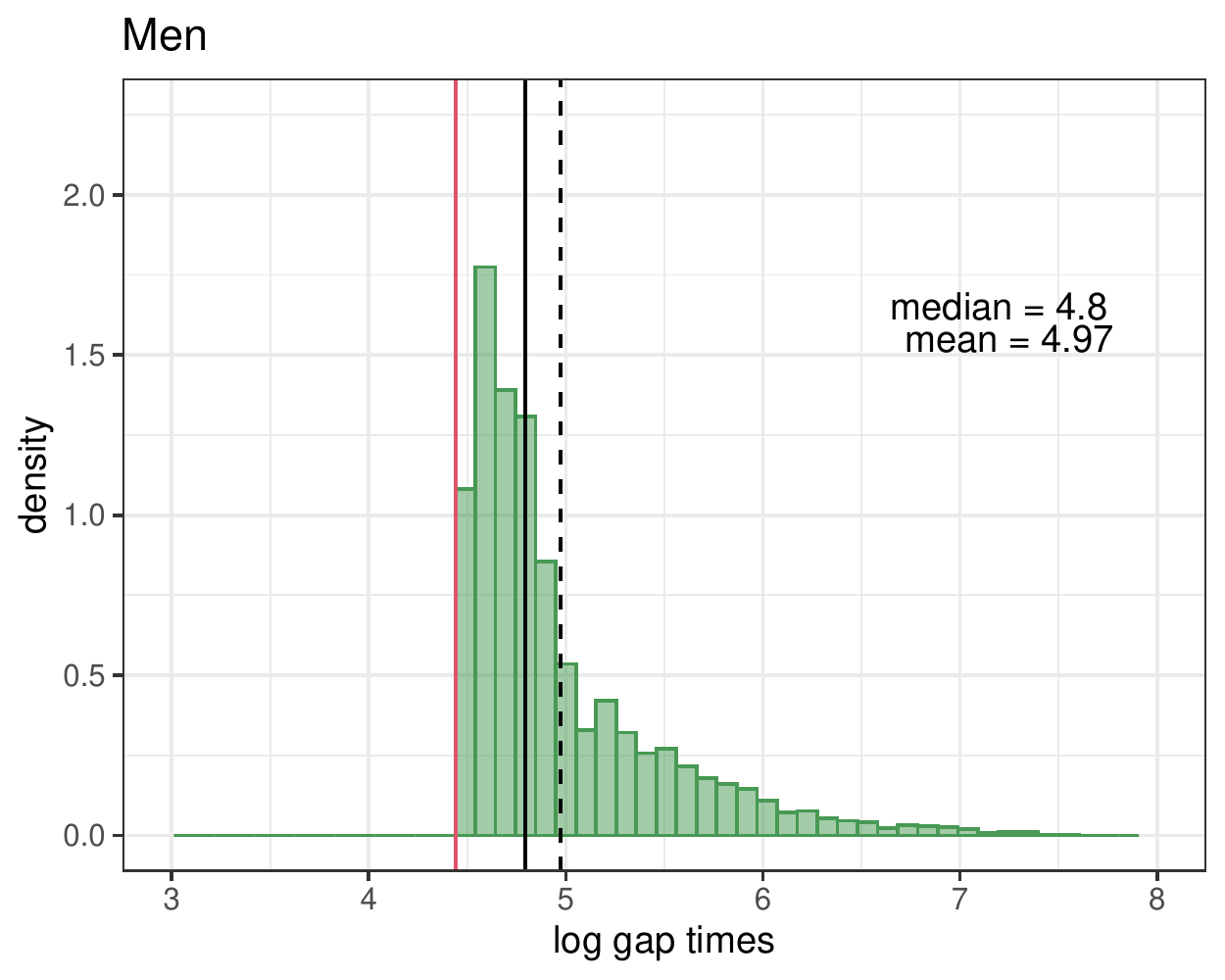}
	\includegraphics[width=0.39\textwidth,height=0.3\textwidth]{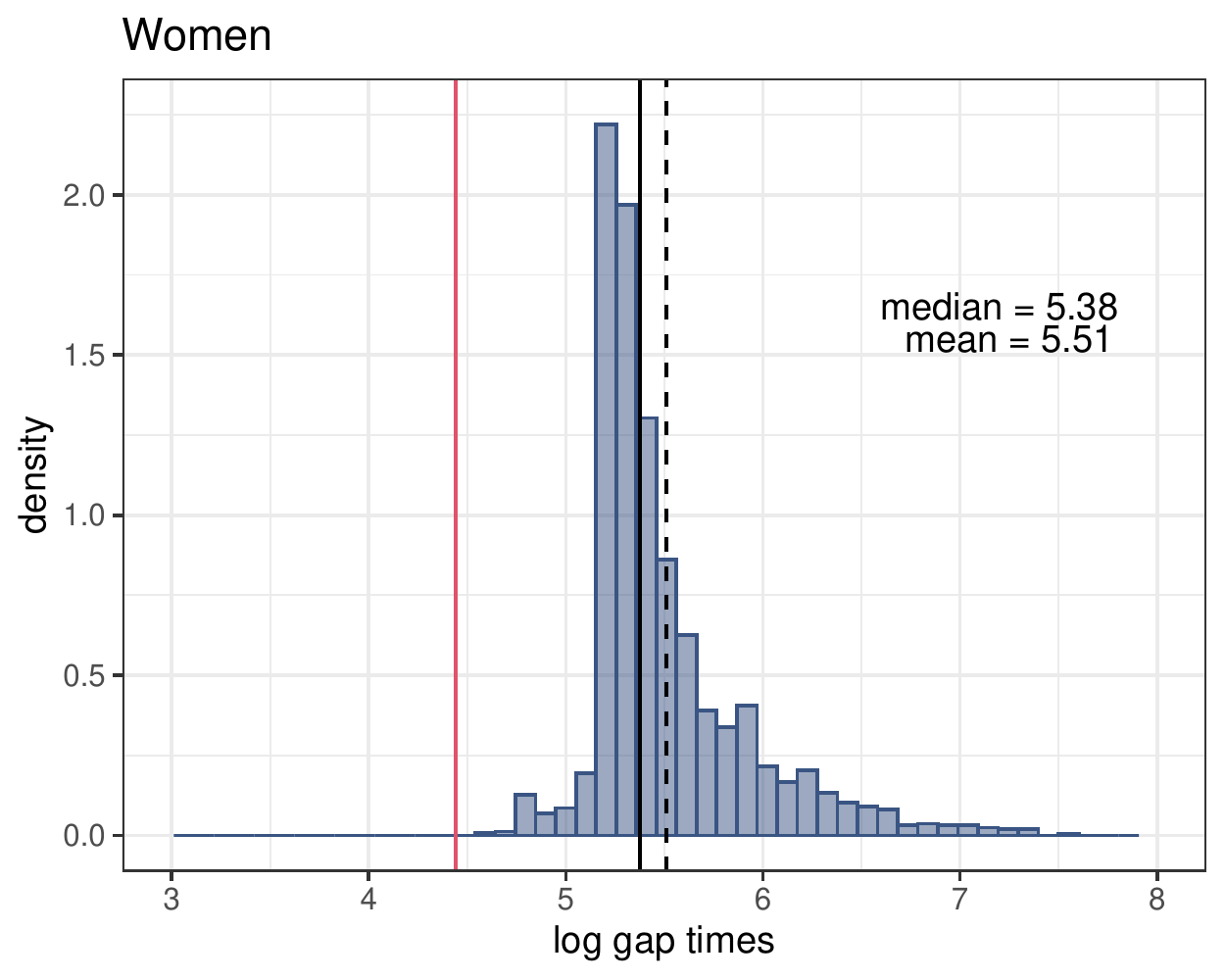}
	\caption{
		Histogram of the logarithm of the observed gap-times grouped by gender; male donors on the left and female donors on the right. The red line denotes the minimum waiting time between two donations, according to the Italian law. The black continuous and dashed lines denotes the empirical median and mean, respectively.
	}
	\label{fig:hist_men_women}
\end{figure}

The statistical focus is the clustering of donors according to the trajectories of gap times. 
Figure~\ref{fig:hist_men_women} reports the histogram of gap times (in the $\log$-scale) for men and women.   
The skewness of these histograms can be explained since, according to the Italian law,
the maximum number of whole blood donations is 4 per year for men and 2 for women, with a minimum of 90 days between a donation and the next one.
Note that the minimum for men is around $4.47$ ($\e^{4.47} \simeq 87$ days), while in median the gap time for the men is $121$ days.
For women, the distribution has a median approximately in $5.24$ in the log scale: this means $189$ days, that corresponds to about 6 months.
Observe that donors may donate before the minimum imposed by law, under good donor's health conditions and the physician's consent.

\begin{figure}[!h]
		\centering
		\includegraphics[width=0.39\textwidth,height=0.3\textwidth]{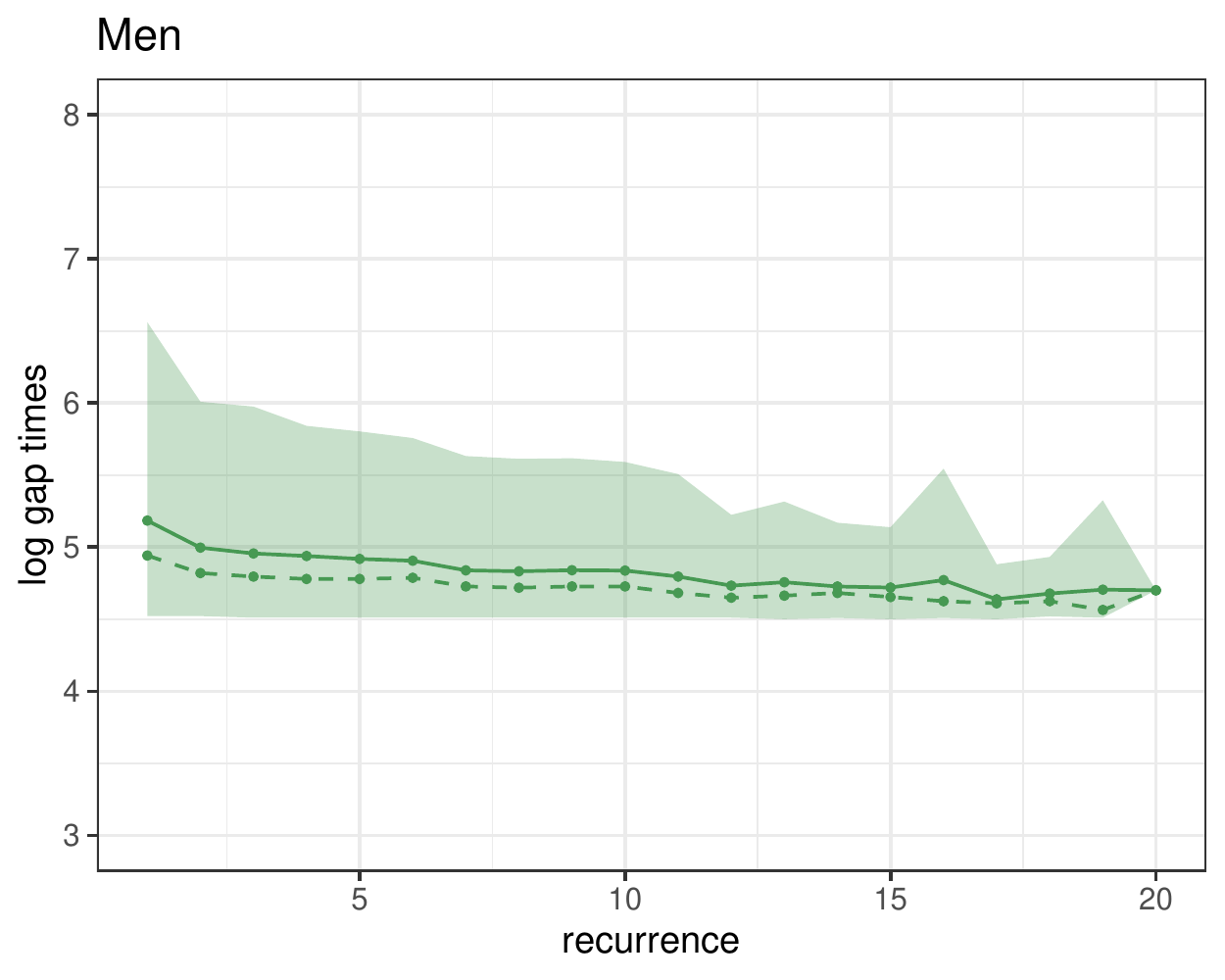}
		\includegraphics[width=0.39\textwidth,height=0.3\textwidth]{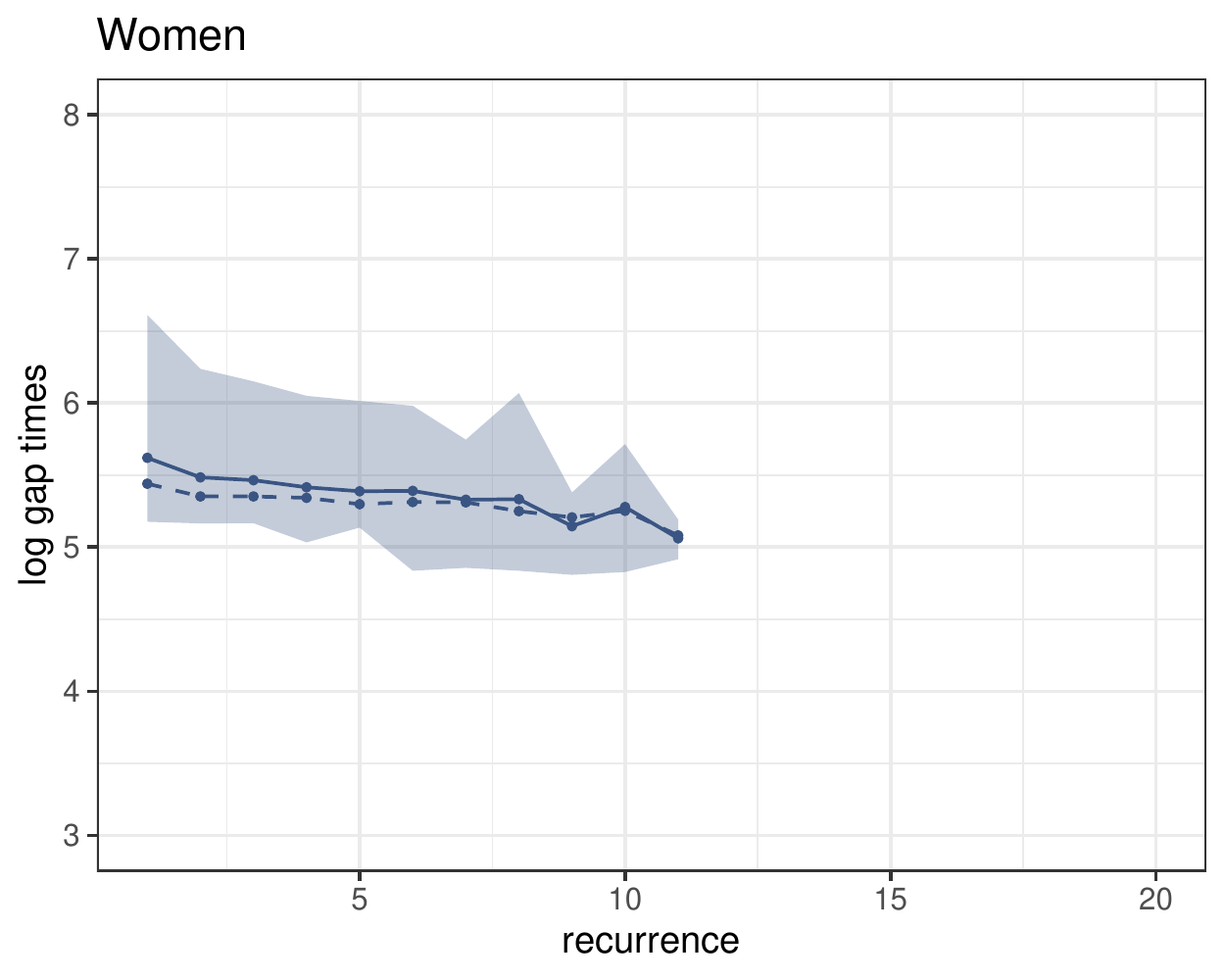}
		\caption{
			Sample mean (continuous line), median (dashed line), and $90\%$ empirical quantile band of the  recurrent gap times, reported on log scale, for each value of $j \in \{1, \dots,20\}$, according to gender men (left) and women (right).
		}
		\label{fig:media_mediana}
\end{figure}

Figure~\ref{fig:media_mediana} reports the mean and median trajectories of gap times
for any recurrence $j=1,\ldots, 20$. 
The average values  decrease as $j$ increases, 
because, as the number of donations increases, the more \textit{loyal} and regular the donor is. 
Note that donors enter  the study randomly in the whole time window. The number of donors for each $j=1,\ldots,20$ is decreasing: there are $2\,912$ donors with at least the first gap time, but only two  with $20$ gap times.

Among different covariates available, we selected some of them which are known to be associated with the gap times, according to a preliminary study \citep[see][]{tesi_gianoli}:
\begin{itemize}
	\item[-] Gender: indicator of gender, 1 if woman, 0 if man;
	\item[-] Blood group: 4-level categorical variable, equal to 0, A, B, and AB;
	\item[-] RH: rhesus factor, 1 if it is positive, 0 if negative;
	\item[-] Smoke: indicator of smoking habit, i.e., 1 if the donor regularly smokes, 0 otherwise;  
	\item[-] Age: age in years at the first donation (at the entrance in the study);  
	\item[-] BMI: body mass index (at the entrance in the study). 
\end{itemize}
Covariates as weight, height and smoke are not directly controlled by AVIS physicians, but are communicated by donors themselves, so that they can be inaccurate.
Table~\ref{tab:freq_static} shows the empirical frequencies for the categorical covariates listed above.
\begin{table}[!h]
	\centering
	\begin{tabular}{ccccccc}
		\hline
		Gender &    \multicolumn{4}{c}{Blood Type}& RH & Smoke \\
		Female & A & B & AB & 0 & + & yes \\
		\hline 
		31.39\% & 38.11\% & 12.33\%&3.91\% &45.64\% & 86.74\% & 32.69\% \\ 
		\hline
	\end{tabular}
	\caption{Empirical frequencies of the static covariates.
	}\label{tab:freq_static}
\end{table}

As far as the age (in years) at the first donation is concerned, 
sample statistics give that the minimum is 18, the maximum is 
a maximum of 68, empirical quantiles of order 25\%, 50\%, 75\% equal to 
27, 35, 44, while the empirical mean and standard deviation are $33.83$ and  $10.27$. 
Similar sample statistics for the BMI values at first donation are 
21.56, 23.93 and 25.70 (sample quartiles) and  $23.93$ and $3.37$ (sample mean and standard deviation).

\subsection{A framework for recurrent events}
\label{sec:rec_events}

Let $n$ be the number of individuals (donors) and $T_{i,t}$ be the time of the $t$-th donation of donor $i$. We assume that $0:=T_{i,0}$ corresponds to the time of first donation for each $i$ 
and that individual $i$ is observed over the time interval $[0,\tau_i]$, where $\tau_i$ denotes the censoring time of the $i$-th observation. If $m_i$ events are observed at times $0<T_{i,1}<\cdots < T_{i, m_i} < \tau_i$, let $W_{i,t}=T_{i,t}-T_{i,t-1}$ for $t=1,\ldots,m_i$ denote the waiting times (gap times) between events of subject $i$ and $W_{i,m_i+1}= T_{i,m_i+1}-T_{i,m_i}$, assuming that  $T_{i,m_i + 1} > \tau_i$ denotes the $(m_i + 1)$-th gap time for the $i$-th donor censored at time $\tau_i$. 
We  assume that the study has been administratively censored, i.e.,  censoring and observations are independent. 
Further, our approach considers the time of all first donations as known. 
We aim at modelling the waiting times $W_{i,t}, t = 1, \dots, m_i$, for $i = 1, \dots, n$, by incorporating some exogenous information in the prior distribution of the latent partition in form of covariates.

Let  $Y_{i,t} = \log (W_{i,t})$ for all $i$ and all $t$, and let ${\bm Y}_i:=(Y_{i,1}, \dots , Y_{i,m_i},Y_{i,m_i+1})$.  We assume that gap times are conditionally independent within clusters , i.e., $f(\bfy^*_j\mid\bfx^*_{j}, \theta_j^*) = \prod_{i\in A_j}f({\bm y}_i|\mathbf{x}_{i}, \theta_j^*)$, 
but differently from Section~\ref{sec:model}, each ${\bm Y}_i$ has dimension $m_i+1$. Furthermore, the evaluation of the sampling distribution includes the information on censoring of the $(m_i+1)$-th gap time for each $i=1,\ldots,n$. For each $t=1,\ldots, m_i+1$ and each $i$ in cluster $A_j$, $j = 1, \dots, k_n$,   we assume that 
\begin{equation}
\label{eq:like_avis}
\begin{cases}
&Y_{i,t}\mid s_i=j, \boldsymbol\beta_0, \boldsymbol\beta_t, \alpha_j,\psi_j, \sigma^2_j, \eta_{i,t} \ind \calN\left(\alpha_j+
\boldsymbol\beta_0^\intercal{\mathbf x}_{i}+\boldsymbol\beta_t^\intercal{\mathbf x}_{i,t}
+\psi_j \eta_{i,t}, \sigma^2_j\right)\\ 
& \eta_{i,t} \iid \mathrm{TN}_{[0, +\infty)}(0,1)
\end{cases}
\end{equation}
 where $\eta_{i,t}$ are latent variables from the standard half-normal distribution and $s_i$ represents the cluster allocation of individual $i$.  
Note that \eqref{eq:like_avis} corresponds to assuming that $Y_{i,t}$ has a skew-normal distribution.  This is motivated by the strong asymmetry in the distribution of the logarithm of the gap-times (see Figure~\ref{fig:hist_men_women}).
Skew normal mixture models have been successfully employed in various contexts; in the Bayesian framework
see, for instance,  
\cite{bayes2007bayesian}, \cite{fruhwirth2010bayesian}, \cite{arellano2007bayesian}, \cite{canale2013informative}. For a definition and its  properties, see \cite{azzalini2005skew} and \cite{arellano2006unification}.
We use parameterization as in \cite{fruhwirth2010bayesian}, which takes into account a skewness parameter, in addition to location and scale parameters. 
%
Note that, in \eqref{eq:like_avis}, the conditional distributions of the gap times on the log scale in cluster $A_j$ share the group-specific parameter $\theta_j^*=(\alpha_j,\psi_j,\sigma^2_j)$, where $\alpha_j$ is the random intercept,   $\psi_j/\sigma_j$ is the skewness parameter \eale  and $\sigma_j$ is a scale parameter.  From \eqref{eq:like_avis}, the expectation of $Y_{i,t}$, in addition to the two linear terms,  is $\alpha_j+\psi_j\sqrt{2/\pi}$, while its variance is $\sigma_j^2+\psi_j^2(1-2/\pi) $.

As far as the linear predictor is concerned, 
we distinguish regression parameters corresponding to fixed-time covariates (${\bm\beta}_0$) from the parameters referring to time-varying covariates (${\bm\beta}_t$), and 
${\mathbf x}_{i} $ includes $p_1$ fixed-time covariates and 
${\mathbf x}_{it}$ denotes $p_2$ time-varying covariates. No intercept is included in the linear predictor to avoid identification issues with the cluster-specific random intercept $\alpha_j$.  
The prior we assume is described as follows
\allowdisplaybreaks
\begin{align}
& \boldsymbol\beta_0 \sim \mathrm{N}_{p_1}\left(\mathbf{0}, \Sigma_0 \right)
\label{eq:prior_beta0}\\
& \boldsymbol\beta_1,\dots,\boldsymbol\beta_{J}\mid \xi^2_1,\dots, \xi^2_{p_2} \stackrel{iid}{\sim} \calN_{p_2}\left( \mathbf{0}, diag(\xi^2_1,\dots, \xi^2_{p_2} )\right), \quad \xi^2_1,\dots, \xi^2_{p_2} \stackrel{iid}{\sim} \mathrm{IG}(\cdot; \nu_0,\gamma_0)\\
& p(\rho_n\mid\mathbf{x}_1, \dots,\mathbf{x}_n) \sim \textrm{PPMx-mixt} \\
&  ( \alpha_j, \psi_j, \sigma^2_j )   \mid \rho_n \stackrel{iid}{\sim} P_0= \ 
\calN_2 \left((\alpha_j, \psi_j)^\intercal;(\alpha_0,\psi_0)^\intercal, \sigma_j^2 \diag(\kappa_0,\kappa_1) \right) 
\times \mathrm{IG}\left(\sigma^2_j;a,b\right). 
\label{eq:prior_P0}  
\end{align}
Note that the number $k_n$ of cluster-specific parameters is determined by $\rho_n$, and its is random. 
Notation $\mathrm{IG}(\cdot;a,b)$ denotes the inverse-gamma density with mean $b/(a-1)$.
Here $\diag(\xi^2_1,\dots, \xi^2_{p_2} )$ is a diagonal matrix which entries $\xi^2_1,\dots, \xi^2_{p_2}$. 
In our specific case, $p_2 = 1$ and the distributions of $\boldsymbol\beta_1, \dots, \boldsymbol\beta_J$ collapse on a univariate Gaussian distribution, where $J=\max_{i}(n_i+1)$ is the maximum number of gap times. 
Notation PPMx-mixt denotes the prior described in Section~\ref{sec:model}.
We assume that the cohesion function $c(u, n_j)$ and the intensity $\rho(\D s)$ are those corresponding to the NGG process. 
Note that  the choice of $P_0$ yields conjugacy of the associated full-conditional \citep[see][]{fruhwirth2010bayesian}.

The same covariates may enter both in the linear predictor and in the prior of the random partition.  
In this application, after covariate selection via LPML (log pseudo marginal likelihood) evaluation, we choose to include Gender, Blood Group, RH, Smoke and BMI (at the first donation) in the linear term, so that $p_1=7$ considering dummy variables too. The only time-varyng covariate included in the linear term is   Age at the $t$-th donation. 
Only static covariates enter in the prior of the random partition: Gender, Blood Group, RH, Smoke, Age at the first donation and BMI at the first donation.

%
%

\subsection{Posterior inference}
\label{sec:post_inf}
To perform posterior inference for model \eqref{eq:like_avis}-\eqref{eq:prior_P0}, we modify the Gibbs sampler in Section~\ref{sec:ppmx_gibbs} of the Appendix to take into account the likelihood for recurrent events. See Section~\ref{sec:avis_gs} of the Appendix for details. 
We fix hyperparameters as follows: the covariance matrix $\Sigma_0$ is assumed equal to  $\diag(1,\dots,1)$ and 
$(\nu_0, \tau_0) = (2, 1)$ so that $\xi^2_1$, as $p_2 = 1$, has prior mean equal to 1 and infinite prior variance; we assume $\alpha_0 = \psi_0=0$ and  $\kappa=0.5$ (see \eqref{eq:NGG_cohesion}). Since $n_j-\sigma$ is the unnormalized weight that a \textit{new} item is assigned to cluster $A_j$, $\sigma$ in \eqref{eq:NGG_cohesion} is a key hyperparameter; hence, we assume three different values for $\sigma$ and report the associated posterior estimates in Table~\ref{tab:avis_tests} for sensitivity analysis. 
 The distance $d(\mathbf x_i, \mathbf x_j)$ entering in the definition of the similarity fuctions is defined in \eqref{eq:sum_of_dist}. 
Every run of the Gibbs sampler produced a final sample size of $10\, 000$ iterations,
after a burn-in of $5\,000$ iterations. In all simulations, convergence was checked using both visual inspection and standard diagnostics.

\begin{table}[!h]
	\centering
	\begin{tabular}{cc|rr|rr}
		\hline
		&&\multicolumn{2}{c|}{$g_C$} &\multicolumn{2}{c}{$g\equiv 1$} \\
		$\lambda$  & $\sigma$ & LPML &$\hat K_{\mathrm{VI}}$ &  LPML &$\hat K_{\mathrm{VI}}$ \\
		\hline
		0.005 & 0.001 & -22\,032.51 & 5 & -22\,424.91 & 5\\    
		0.010 & 0.001 & -22\,068.25 & 5 &&\\    
		0.100 & 0.001 & -21\,812.24 & 6 && \\    
		\hline
		0.005 & 0.150 & -21\,990.26 & 6   & -22\,252.37 & 6\\    
		0.010 & 0.150 & -21\,758.68 & 5   &&\\    
		0.100 & 0.150 & \cellcolor{lgray} -21\,221.55 & \cellcolor{lgray} 5 &&  \\   
		\hline
		0.005 & 0.300 & -22\,150.35 & 5 & -22\,311.26 & 7\\    
		0.010 & 0.300 & -22\,012.24 & 6 &&\\    
		0.100 & 0.300 & -21\,672.05 & 7 &&  \\   
		\hline
	\end{tabular}
	\caption{LPML and number of clusters in the estimated partition, obtained minimizing a
 posteriori VI, for the blood donation data, for different values of $\lambda$  and  $\sigma$ and similarities $g_C$, $g\equiv 1$.  In evidence: the best model in terms of LPML.}
	\label{tab:avis_tests}
\end{table}
Table~\ref{tab:avis_tests} shows the number of clusters of the estimated partition and LPML \citep{Chr10} values with similarity functions $g_C$  and $g\equiv 1$ (no covariates in the prior), for different values of the \textit{temperature} hyperparameter $\lambda$ and the \textit{reinforcement} parameter $\sigma$. By its definition, the larger the LPML is, the better the model fits the data.  
There is a clear effect of $\sigma$, $\lambda$ and covariates (through $g_C$) on LPML: when $\lambda$ or $\sigma$ increase,  LPML decreases, corresponding to a better fit of the model. Values of LPML for $g_C$ are much larger than in the case of $g\equiv 1$. For any of the hyperparameter values in the table, 
we have computed an estimate of the random partition for the sample donors, 
 minimizing a posteriori the expectation of  the variation of information (VI) loss function (see, for instance, \citet{Wad18}). We report the number $\hat K_{\mathrm{VI}}$ of estimated clusters  in Table~\ref{tab:avis_tests}.   Since the cardinality of the visited partitions is quite large, 
as suggested by \citet{Wad18},
from the MCMC estimate of the posterior similarity matrix, 
 we consider all the partitions designed  by a hierarchical clustering algorithm with complete linkage.   
Then, as the point estimate, we select the partition that achieves the minimum value of the posterior loss function. 
 It is clear that $\hat K_{\mathrm{VI}}$ is robust with respect to  the effect of covariates in the prior and changes in 
$\sigma$ and $\lambda$, though as expected,  $\hat K_{\mathrm{VI}}$ increases with $\sigma$. 
This is an aspect of the well-known trade-off between estimation of the number of clusters and posterior predictive checks, especially in case of misspecified models \citep[see, for instance][Section 7]{beraha2022mcmc}, which generally improve with a large number of clusters.  
 
The best model fit, in terms of LPML, is obtained when $\lambda = 0.100$ and $\sigma = 0.150$.
The rest of posterior inference that we report here below is computed for these values of the hyperparameters.   
Note that $\sigma=0.001$ in Table~\ref{tab:avis_tests} 
approximates the cohesion function yielded by the Dirichlet process as in \cite{PPMx_JSPI10} and \cite{PPMx_JCGS11} (though they use a different similarity). 

\begin{table}[!h]
	\centering
	\begin{tabular}{l|rcc}
	\hline
		Covariate & Median & $95\%$ C.I. & $\max\{\Pr(\beta_j > 0), \Pr(\beta_j < 0)\}$\\
		\hline
		BMI & -0.060 & (-0.077;-0.043) & 1.000\\
		Gender & 1.119 & (0.938;1.329) & 1.000 \\
		Blood Type 0 & 1.137 & (0.779;1.480) & 1.000 \\
		Blood Type A & 1.131 & (0.768;1.486) & 1.000 \\
		Blood Type B & 1.230 & (0.755;1.692) & 1.000 \\
		RH & 0.533 & (0.295;0.755) & 1.000 \\
		Smoke & 0.339 & (0.148;0.526) & 0.999 \\
		\hline
	\end{tabular}
	\caption{Posterior summaries of the fixed-time regression coefficients $\bm \beta_0$ for the blood donation application.}\label{tab:avis_coefs}
\end{table}

Table~\ref{tab:avis_coefs} shows  posterior means of the regression coefficients of the fixed-time covariates. All the fixed-time covariates included in the study are  significantly different from zero; see the last column. 
The average of log-gap time increases for donors with blood group 0, A and B with respect to the reference level AB.
 Of course, women exhibit longer gap-times in accordance with 
	the Italian law.
\begin{figure}[!htb]
	\centering
	\includegraphics[width=0.99\textwidth]{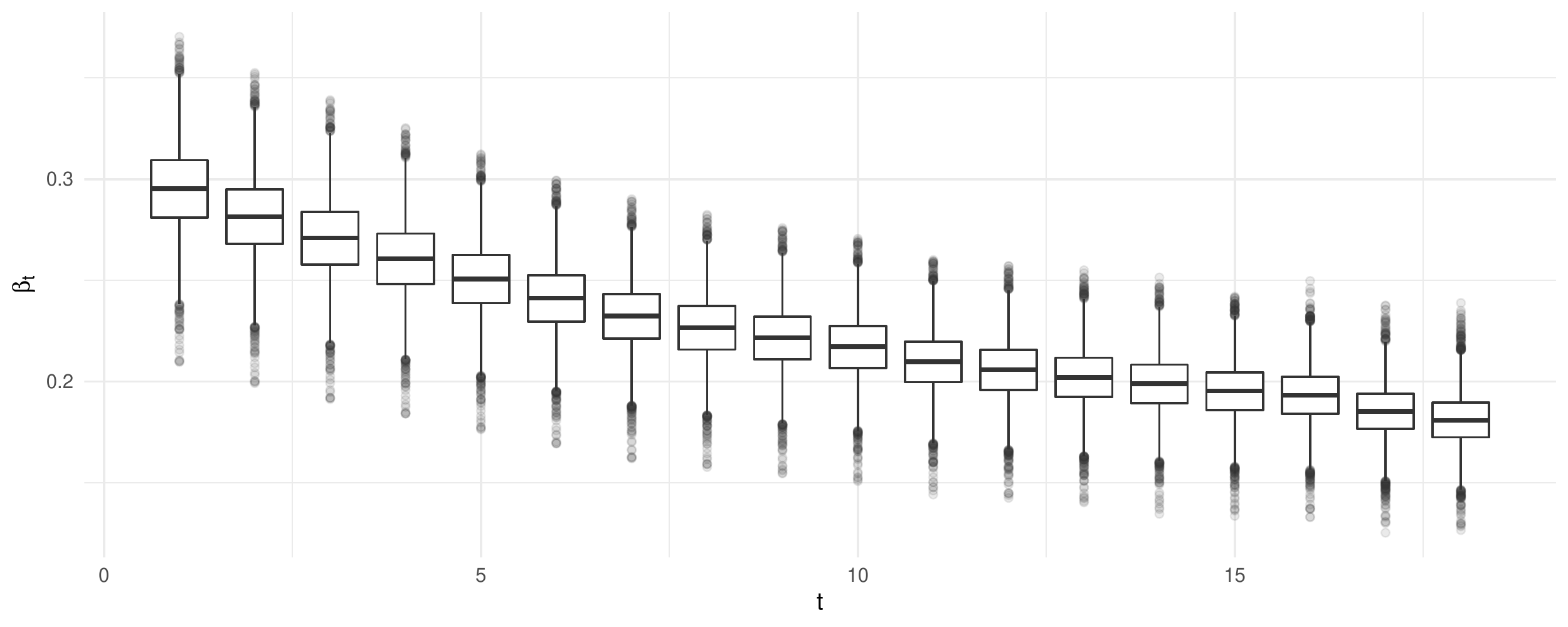}
	\caption{Recurrent gap times (on the log scale) by estimated cluster for the blood donation application. For each cluster we draw the sample mean (continuous line), the sample median (dashed line) and 90\% sample quatile band in each cluster.  The black continuous and dashed lines denote the overall mean and median, respectively, while the number $n$ denotes the cluster size. 
	}
	\label{fig:avis_time_coeff}
\end{figure}
Figure~\ref{fig:avis_time_coeff} shows the regression coefficients for the only time-dependent covariate included in the study (age of the donor). These parameters are significantly different from zero for all donation occasions. Further, as the occasion of donation increases, the impact of the age on the log-gap time decreases in magnitude, 
implying  that loyal donors (i.e., donors who have donated many times) are  less subject to age differences. 


\begin{figure}[!htb]
		\centering
		\includegraphics[width=0.99\textwidth]{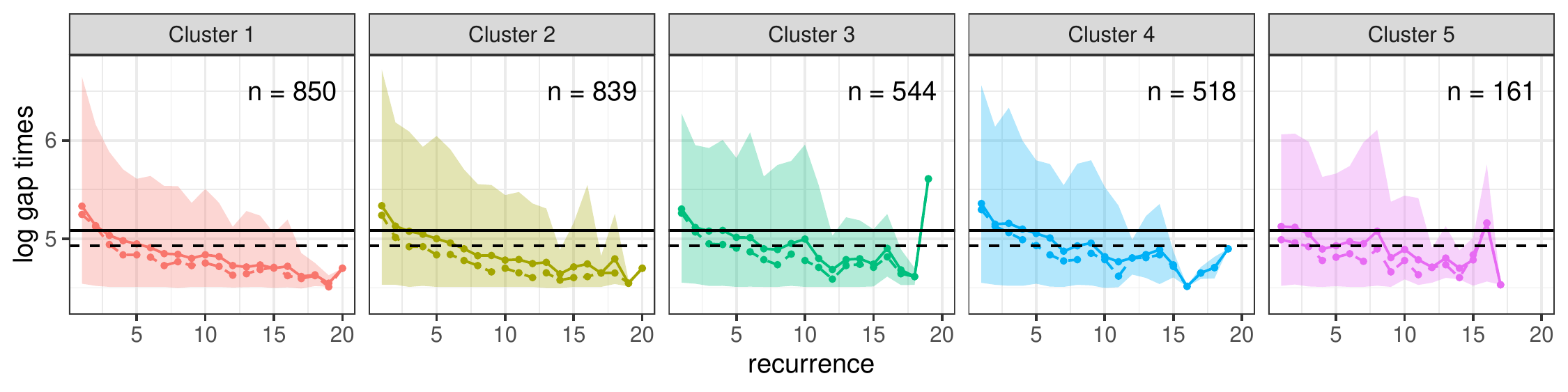}
		\caption{Recurrent gap times (on the log scale) by estimated cluster for the blood donation application. For each cluster we draw the sample mean (continuous line), the sample median (dashed line) and 90\% sample quatile band in each cluster.  The black continuous and dashed lines denote the overall mean and median, respectively.}
		\label{fig:clusters}
\end{figure}

Figure~\ref{fig:clusters} shows the trajectories of the observed log-gap times grouped by the estimated clusters as explained before.  It is clear from the cluster sizes that   \textit{rich-get-richer} property of the cohesion associated to the Dirichlet process is here mitigated. We do not observe substantial differences among the log-gap times in the estimated clusters, but Cluster 1 seems to groups longer trajectories  (see also the last column of Table~\ref{tab:group_summaries}).

\begin{table}[!h]
	\centering      
	{\small    
	\begin{tabular}{l|rrrrrrrrrrrrrrr}
		\hline
		 &Age &BMI&Gender &    \multicolumn{4}{c}{Blood Type}& RH & Smoke & No. Donations\\
		&&&Female & A & B & AB & 0 & + & yes & mean \ (sd)\\
		\hline 
		Cluster 1 &46.81 &24.48 &36.35\% &40.29\% & 12.51\% & 3.34\% & 43.86\% & 88.92\% & 32.78\% & 4.24  (3.63)\\ 
		Cluster 2 &33.92 &24.11 &29.06\% &38.59\% & 11.88\% & 3.18\% & 46.35\% & 87.18\% & 37.53\% & 3.92  (3.51)\\ 
		Cluster 3 &28.16 &23.77 &33.64\% &34.01\% & 12.13\% & 3.49\% & 50.37\% & 87.13\% & 31.07\% & 4.08  (3.39)\\ 
		Cluster 4 &22.83 &22.95 &32.24\% &38.22\% & 12.16\% & 5.02\% & 44.59\% & 84.94\% & 27.80\% & 3.23  (2.82)\\ 
		Cluster 5 &20.26 &23.79 & 7.45\% &37.89\% & 14.91\% & 8.70\% & 38.51\% & 77.64\% & 27.95\% & 4.49  (3.10)\\ 
		\hline
		All &33.83 &23.93 &31.39\% & 38.11\% & 12.33\%&3.91\% &45.64\% & 86.74\% & 32.69\% & 3.95 (3.40)\\ 
		\hline
	\end{tabular}
	}
	\caption{Empirical summaries of the covariates within each estimated cluster for the blood donation application.
	}\label{tab:group_summaries}
\end{table}

Table~\ref{tab:group_summaries} reports empirical summaries of the covariates (included in the prior) within each estimated cluster.  We report empirical means for any continuous covariate and empirical frequency for the binary or categorical covariates. The last column display the empirical average and standard deviation for the number of recurrences ($m_i$'s) per cluster. 
Cluster~1 groups older donors, since the cluster mean is one standard deviation above the overall mean. These donors also have a slight higher BMI. Cluster~2 contains donors with age at the first donation that is close to the overall mean (33.83 years), while Cluster~3 groups younger donors than in Cluster 2, with a lower percentage of smokers. Clusters~4 and 5 group very young donors, and Cluster 5 is mostly made of men. 


To study if the inclusion of $g_C$ in the prior affect the posterior predictive inference, 
we consider a cross-validation approach where we subsample $50$ different training subsets  containing  $90\%$ of the donors. For each training subset we compute the associated posterior and use the remaining donors as testing set for prediction.  We adopt the same model specification given in Section~\ref{sec:post_inf} with $\lambda=0.1$ and $\sigma=0.15$. Posterior predictive inference has been computed as explained at the end of Section~\ref{sec:posterior_calc}.  
We also compare with the case $g \equiv 1$. 
For each training subsample, we ran the Gibbs sampler for $3\, 000$ iterations,
of which $2\,000$ were discarded as burn-in iterations. 
%
The root mean square errors (rMSE) between observed and predicted log-gap times are $1.825$ and $2.242$ for the PPMx-mixt with $g_C$ and the model with $g \equiv 1$ respectively, i.e., the rMSE is decreasing by $18.6\%$ while including the covariates in the partitions' distribution through the similarity function $g_C$.

\section{Discussion}\label{sec:discussion}

In this work we propose a regression model for gap times of recurrent events, where parameterization includes the partition $\rho_n$ of the blood donors. The prior we assume includes covariate information, encouraging two individuals to be co-clustered if they have similar covariate values. 
We have seen that including covariate information improves the posterior predictive performance and helps interpret the estimated clusters in terms of covariates. 
Thanks to the introduction of a latent variable $u>0$, we are able to express the cohesion function in the prior, and hence the whole prior for the random partition of the sample, as mixtures of PPMx. 
Our new prior on the random partition is given via cohesion and similarity functions, as specified by PPMx models \citep{PPMx_JCGS11}.  

Differently from \cite{PPMx_JCGS11}, we assume a wide class of  non-increasing similarity function of the cluster compactness which takes values in $(0,1]$. We propose three examples of such non-increasing functions, emphasizing their properties and their effects on the posterior predictive distribution of the model.
Cross-validated posterior predictive root mean-squared errors for the AVIS dataset shows that the inclusion of the similarity function $g$ in the prior for the random partition yields a lower value than  in the case with no covariates  in the prior.  
For our motivating application, we have introduced a model for recurrent data when gap times assume skew distributions. The model also includes cluster specific random effects modelled via PPMx-mixt. We estimate five clusters of homogeneous donors. This grouping helps in identifying characteristic and important features (covariates) of individuals resulting in a more accurate prediction of the time of a future donation.   

An interesting characteristic of our model is that, though it clusters donor gap times trajectories, when we aim at interpreting the estimated clusters in terms
of number of distinct gap times trajectories in the responses, we should also consider that the
prior we assume for the random partition of the sample subjects is covariate-dependent. In
fact, some of the estimated clusters are similar when looking at the response trajectories
but different when looking at the covariates.
We consider this aspect as an advantage of all models with covariate-dependent prior for the random partition), included ours, that
allows for greater  flexibility  for clustering, rather than an inadequacy.



As far as the sensitivity of the model to the distance $d$ is concerned, different choices can be considered and combined with the functions $g_A$, $g_B$ and $g_C$ that we used in the manuscript, but not all the distances support the increasing properties described in Section~\ref{app:secA} of the Appendix. The similarity functions $g$ that we propose must be calibrated via parameter $\lambda$ and we discuss how we can fix it. 
This is a key parameter that prevents the overpowering effect of covariates on clusters  with respect to likelihood. For a thorough discussion on the  calibration of similarity functions in PPMx models, see 
\cite{page2018calibrating}. 


The  pitfall  of our strategy consists in  its computational cost. Future   work   may consider the use of approximate sampling strategies to overcome this limitation.   Model flexibility can be enhanced by assuming  some of the hyperparameters of the model, such as those in the cohesion  and the similarity functions, to be random.  However,  this extension is certainly not trivial, due to the intractability of the normalization constant $M_g(\bfx_1, \dots, \bfx_n)$ in  the prior distribution of the random  partition.

\section*{Acknowledgments}
 The authors are grateful to Dr. Ilaria Bianchini for her contribution to the construction of the model and the computational strategy at an early stage of this manuscript. Her work has been part of her PhD thesis.
 Raffaele Argiento was partially supported by grant \virgolette{CLUstering: Bayesian Partition Models for Precise
	Medicine (CluB: PMx2)}, funded by \emph{Fondo di Beneficienza di Intesa Sanpaolo (Italy)}.
The authors also thank Dr. Sergio Casartelli, general manager of AVIS Milano, for kindly providing data and support for the interpretation of the posterior inference. 

\bibliography{bibliography.bib}

\appendix

\section{$\mathcal D_{A_j}$ is decreasing with the size of $A_j$}\label{app:secA}

 Let $A_j$ be an element of the partition  of the sample labels. Since $\mathcal{D}_{A_j} =  \sum_{i \in A_j}d(\mathbf{x}_i, \mathbf{c}_{A_j})$ and the  Fr\'echet mean of order one is defined as
\begin{equation*}
\label{eq:frech}
\mathbf c_{A_j} = \underset{\mathbf c \in \mathbb X}{\arg\min} \left\{ \sum_{i \in A_j} d(\mathbf x_i, \mathbf c) \right\},
\end{equation*}
it is easy to check that 
\begin{equation}\label{eq:inc_property}
\begin{split}
\mathcal{D}_{A_j\cup \{\ell\}} &=  \sum_{i \in A_j\cup \{\ell\}}d(\mathbf{x}_i, \mathbf{c}_{A_j\cup \{\ell\}}) = \sum_{i \in A_j}d(\mathbf{x}_i, \mathbf{c}_{A_j\cup \{\ell\}}) + d(\mathbf x_\ell, \mathbf{c}_{A_j\cup \{\ell\}})\\
&\geq \sum_{i \in A_j}d(\mathbf{x}_i, \mathbf{c}_{A_j\cup \{\ell\}}) \geq \sum_{i \in A_j}d(\mathbf{x}_i, \mathbf{c}_{A_j}) = \mathcal{D}_{A_j} .
\end{split}
\end{equation}

\section{Supplementary figure}\label{app:secB}

\begin{figure}[h]
	\includegraphics[width = 0.99\textwidth]{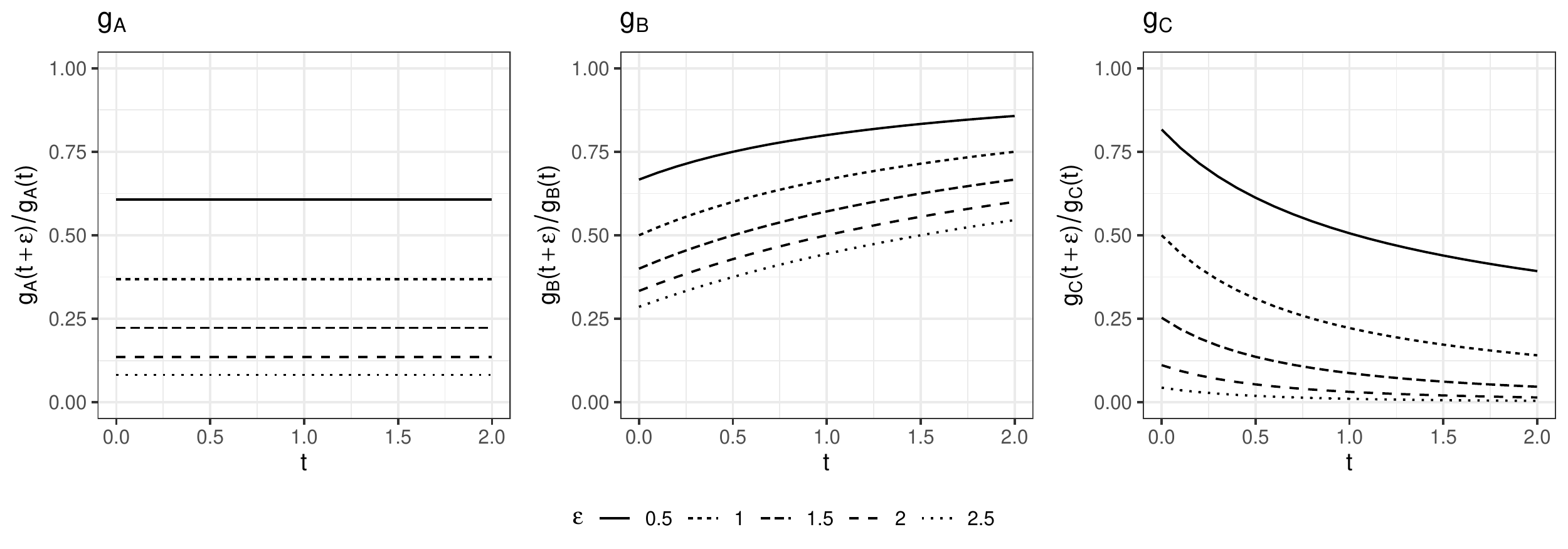}
	\caption{Ratios $g(t + \varepsilon) / g(t)$ as function of $t\in (0, 2)$, for different values of $\varepsilon\in \{ 0.5, 1, 1.5, 2, 2.5 \}$. Left panel: $g_A(\cdot)$. Middle panel: $g_B(\cdot)$. Right panel: $g_C(\cdot)$.}\label{fig:ratios}
\end{figure}

Let $t=\mathcal{D}_{A_j}$ and $t+\varepsilon=\mathcal{D}_{A_j  \cup \{ i\}} $, where $\varepsilon$ represents the increment of the average center-based distance when $\{ x_i\}$ is assigned to cluster $A_j$. Figure \ref{fig:ratios} shows the ratio $g(t+\varepsilon)/g(t)$, for different values of $\varepsilon$ and  similarity functions $g_A$, $g_B$ and $g_C$. See Section~\ref{sec:similarity_g} in the manuscript.

\section{Gibbs sampler for the Gaussian kernel with the linear regressor}
\label{sec:ppmx_gibbs}
In this section, we illustrate the Gibbs sampler P\'olya urn scheme for
model \eqref{eq:lik1}-\eqref{eq:rho_x} in the context of linear regression, i.e. when $f(\bfy_j\mid\bfx^*_j, \bm{\theta}_j^*)=\prod_{i\in A_j} \phi(y_i;\mathbf{x}_i^\intercal \bm{\beta}^{*}_{ j},\sigma^{ 2*}_{j})$. In this case the cluster specific parameter $\bm\theta=(\bm \beta,\sigma^2)$  consists of  the vector of regression  coefficients $\bm \beta$ and the  residual variance  $\sigma^2$. We assume that the base distribution $P_0$ is conjugate to the mixture kernel, that is
$$   \theta=  \left(\boldsymbol\beta, \sigma^2\right) \sim P_0(d\boldsymbol\beta,d \sigma^2)= \mathrm{N}_p(d\boldsymbol\beta; \boldsymbol\mu_0, \sigma^2 B_0 ) \times \mathrm{IG}(d\sigma^2; a_0, b_0).$$
The algorithm  here described  is an extension of Algorithm 8 in \cite{Neal00}, later generalized to normalized completely 
random measures in \cite{LijPru_beyond} and \cite{FavTeh13}. 
It consists of  a marginal MCMC sampler in its simplest form, thanks to the above mentioned conjugacy.
In this case, the  vector of  cluster parameters $ \bm{\theta}^*=(\theta_1^*,\ldots,\theta^*_{k_n})$ can be efficiently marginalized out from the joint distribution \eqref{eq:joint}
obtaining
\begin{equation*}
\mathcal{L}\left(\{y_i\}_{i=1}^n, \rho_n, u \mid  \left\{\mathbf{x}_i\right\}_{i=1}^n \right) \propto D(u,n) \prod_{j=1}^{k_n}  m(\bfy_j) c(u,n_j) g(\bfx^*_j)
\end{equation*}
where 
\[
m(\bfy_j) := \int_\Theta f(\mathbf{y}_j^*\mid \mathbf{x}_j^*, \mathbf{\theta}_j^*)P_0(\D \mathbf{\theta}_j^*)
\] 
is the marginal distribution of data in the $j$-th cluster $A_j$.
Therefore, the Gibbs sampler is obtained by repeatedly sampling from the full-conditionals below.
In what follows,  $\mathcal{L}(\cdot \mid-)$ indicates that we consider the law of the variable $\cdot$ given all remaining variables $-$ (including the data). 
 
\begin{enumerate}
	\item[{[1]}]   Update $u$:   given the partition $\rho_n$, the auxiliary variable $u$ and data are independent, so that 
	$\mathcal{L}(u\mid-)$ $\propto$ $u^{n-1} \e^{-\Psi(u)} \prod_{j=1}^{k_n} c(u,n_j)$ with $\Psi(u) = \kappa \int_0^{+\infty} (1-\e^{-us}) \rho(s) ds$;  in the case 
	of the NGG process, this full-conditional simplifies to
	$$ \mathcal{L}(\D u\mid-) \propto \dfrac{u^{n-1}}{(u+1)^{n-\sigma k_n }}\e^{- \frac{\kappa}{\sigma}\left( \left( u+1\right)^{\sigma}-1\right)} \uno_{(0,+\infty)}(u)\D u.  $$
	In particular, to sample from this distribution, we
	use a simple Metropolis-Hastings update with a Gaussian proposal kernel truncated in $\left(0, +\infty\right)$.

	\item[{[2]}] Update the group specific parameters: 
	each $\theta_j^* = \left( \boldsymbol\beta_j^*,  \sigma_j^{2*}\right)$, for $j=1,\dots,k_n$ is updated within each cluster according to the usual parametric 
	update in the conjugate case with normal likelihood and normal-inverse gamma prior distribution. In particular, we have that, for each $j=1,\dots, k_n$, the cluster specific parameters can be sampled independently from the following distributions:
	$$\boldsymbol\beta_j^ *\mid \sigma_j^{2*}, - \ind \mathrm{N}_p\left(\bm\mu_j ,  \sigma_j^{2*}  B_j\right)$$
	where $B_j = \left(B_0^{-1} + \sum_{i \in A_j} \mathbf{x}_i \mathbf{x}_i^\intercal \right)^{-1} $, $\bm\mu_j =  B_j\left(B_0^{-1}\bm\mu_0 + \sum_{i \in A_j} y_i \mathbf{x}_i \right)$
	and
	$ \sigma_j^{2*}\mid- \sim \mathrm{IG}\left(a_j, b_j\right),$
	with $a_j =a_0 + \dfrac{n_j}{2}$ and $b_j = b_0 + \dfrac{1}{2}\left(\boldsymbol\beta_0^\intercal B_0^{-1}\boldsymbol\beta_0 
	+ \sum_{i \in A_j}y_i^2 - \boldsymbol\mu_j^\intercal B_j^{-1}\boldsymbol\mu_j \right)$. Here $n_j:=\lvert A_j\rvert$ is the size of cluster $A_j$ in the partition. 
	
	\item[{[3]}] Update the latent partition: the random partition $\rho_n$ is updated using a Gibbs sampling step where the cluster assignment of one item $Y_i$ is updated once at a time. We denote by $\rho_{n-1}^{(-i)}$ the partition of $n-1$ items where the $i$-th item has been removed and by $s_i=j$ the event that $Y_i$ is assigned to cluster $j$, where $j$ varies 
	in $\left\{1, \dots,k_{n-1}^{(-i)},k_{n-1}^{(-i)}+1\right\} $ and  $k_{n-1}^{(-i)}$ is the number of clusters available in the partition without $i$.
	Note that $k_{n-1}^{(-i)}+1$ is included to consider the case where the item forms a new cluster.
	Therefore, we have to sequentially sample from the following conditional distribution, for $i=1,\dots,n$,
\begin{equation}
	\mathcal L \left( s_i =j \mid u,  {\mathbf{x}}, \left\{ y_i \right\}_{i=1}^n, \rho_{n-1}^{(-i)} \right) = 
	\frac{\mathcal{L}\left(\left\{y_i\right\}_{i=1}^n \mid u, {\mathbf{x}}, \rho_{n-1}^{(-i)}, s_i=j \right) 
	   \mathcal{L}\left(s_i=j , \rho_{n-1}^{(-i)} \mid u,  {\mathbf{x}}\right)
	}{
	\mathcal{L}\left(\left\{y_i\right\}_{i=1}^n\mid u, {\mathbf{x}},  \rho_{n-1}^{(-i)}\right)
	\mathcal{L}\left(\rho_{n-1}^{(-i)}	\mid u, {\mathbf{x}} \right)},  
	\label{eq:newclust}
\end{equation}
where $j=1,\ldots, k_{n-1}^{(-i)}+1$, and $\mathbf{x}:=\{\mathbf{x}_i\}_{i=1}^n$.
	Moreover, observe that, for any $l=1,\dots, k_{n-1}^{(-i)}$, the prior on the partition can be written as:
	\begin{align*}
\mathcal{L}\left(\rho_{n-1}^{(-i)}, s_i=j \mid  u,\mathbf{x}\right) &\propto D(u,n) \prod_{l=1}^{k_n} \left(  c(u,n_l) g(\bfx^*_l)\right)\\
	& \propto  D(u,n) \prod_{l=1}^{k_{n-1}^{(-i)}} \left( c(u,n_l) g(\bfx^*_l) \right) c(u,n_j+1) g(\bfx^*_j \cup \{ x_i\})\\
	& =  D(u,n) \prod_{l=1}^{k_{n-1}^{(-i)}} \left( c(u,n_l) g(\bfx^*_l) \right)  \frac{c(u,n_j+1) g(\bfx^*_j \cup \{ x_i\}) }{ c(u,n_j) g(\bfx^*_j )} \\
	& \propto \mathcal L \left(\rho_{n-1}^{(-i)}\mid u,\mathbf{x}\right)  \frac{c(u, n_j+1) g(\bfx^*_j \cup \{ x_i\}) }{ c(u,n_j) g(\bfx^*_j )}
	\end{align*}
	while,  since   $g(\emptyset)=1$, the conditional probability of assigning item $i$ to a new cluster is equal to 
	\begin{align*}
		\mathcal{L}\left(\rho_{n-1}^{(-i)}, s_i= {k_{n-1}^{(-i)}+1} \mid  u,\mathbf{x}\right) &\propto \mathcal L \left(\rho_{n-1}^{(-i)}\mid u,\mathbf{x}\right)  c(u, 1) 
	\end{align*}
The contribution of the likelihood in \eqref{eq:newclust}   can be written as  
	\begin{align*}
	\mathcal{L}\left(\{y_i\}_{i=1}^n\mid u,  {\mathbf{x}}, \rho_{n-1}^{(-i)}, s_i=j\right)&=  \prod_{l=1,l\neq j}^{k_{n-1}^{(-i)}} m(\bfy^*_l )  
	m\left(\bfy^*_j \cup\{ y_i\}\right) \frac{ m(\bfy^*_j ) }{ m(\bfy^*_j) } \\
	&= \mathcal{L}\left(\{y_\ell\}_{\ell\neq i}\mid  \rho_{n-1}^{(-i)} \right) 
	\frac{  m(\bfy^*_j \cup\{ y_i\})}{m(\bfy^*_j) },
	\end{align*}
where $m(\emptyset)=1$ in the case of a new cluster.
	Therefore, \eqref{eq:newclust} becomes

	\begin{align}
	\Pr&\left(s_i=j\mid u,  {\mathbf{x}}, \left\{y_i\right\}_{i=1}^n, \rho_{n-1}^{(-i)} \right) \propto \frac{  m(\bfy^*_j \cup\{ y_i\})}{m(\bfy^*_j) }   \frac{c(u,n_j+1) g(\bfx^*_j \cup \{ x_i\}) }{ c(u,n_j) g(\bfx^*_j )}, \quad j=1,\ldots, k_{n-1}^{(-i)} 
	\label{eq:clusterassign_Gibbs}
\end{align}
and, similarly,
\[
	\Pr\left(s_i=k_{n-1}^{(-i)}+1 \mid u,  {\mathbf{x}}, \left\{y_i\right\}_{i=1}^n, \rho_{n-1}^{(-i)} \right)\propto m( y_i) c(u, 1).
	\]

We sample each $s_i$ is sequentially assigned according to this law. 
	Observe that, because of conjugacy, $m(\bfy^*_j)$ is available analitically and it equals to a Student's $t$ density. 
	
\end{enumerate}


\section{Gibbs sampler for the blood donations application}\label{sec:avis_gs}
In this section we   describe a Gibbs sampler for the posterior of model \eqref{eq:like_avis}-\eqref{eq:prior_P0}.
\allowdisplaybreaks
The state  of the Markov chain is   
$$\left\{ \left\{\eta_{it} \right\}_{t=1}^{m_i+1}, i=1,..,n; \left\{Y_{i ,n_i+1}^{cens} \right\}_{i=1}^n; 
\boldsymbol\beta_0;\left\{\boldsymbol\beta_t\right\}_{t=1}^{J}; \left\{\tau^2_i\right\}_{i=1}^{p_2}; \rho_n; \left\{\left(\alpha_j, \psi_j,\sigma^2_j\right)^\intercal\right\}_{j=1}^{ {k_n}}\right\}.
$$
The full-conditionals are outlined below: we provide the details of the
computation only when the conditional posterior distribution is not straightforward. 
As before,  $\mathcal{L}(\cdot \mid-)$ indicates that we consider the law of the variable $\cdot$ given all remaining variables $-$ (including the data).  
\begin{enumerate}
	\item[{[1]}] Update the latent variables  $\eta_{i,t}$'s: each $\eta_{i,t}$, conditionally on $s_i=j$, is independently sampled from  
	$$\mathcal{L}(\eta_{i,t}\mid-) \propto 
	\exp\left\{-\dfrac{1}{2\sigma^2_j}\left( y_{i,t} - \left(\alpha_j +\boldsymbol\beta_0^\intercal\mathbf{x}_{i}+
	\boldsymbol\beta_t^\intercal \mathbf{x}_{i,t} +
	\psi_j \eta_{i,t}\right)\right)^2 -\dfrac{1}{2}\eta_{i,t}^2 \right\}\mathbb{I}\left(\eta_{i,t}>0\right) $$ 
	which turns out to be a truncated normal, namely
	$$  \eta_{i,t}\mid - \sim \mathrm{TN}_{[0, \infty)}\left( \dfrac{\psi_j}{\sigma^2_j + \psi_j^2} \left( y_{i,t} - (\alpha_j +\boldsymbol\beta_0^\intercal\mathbf{x}_{i}+
	\boldsymbol\beta_t^\intercal \mathbf{x}_{i,t}
	) \right),  \dfrac{\sigma^2_j}{\sigma^2_j + \psi_j^2}\right) $$
	independently for  each $t =1,\dots, m_i+1$ and $i=1,\dots, n$
	such that $s_i=j$.
	
	\item[{[2]}] Update the censored values: the censored observations are independently sampled from
	$$ Y_{i,m_i+1}^{cens}\mid - \sim \mathrm{TN}_{[y_{i,m_i+1}, +\infty)}\left(\alpha_j 
	+\boldsymbol\beta_0^\intercal\mathbf{x}_{i}+
	\boldsymbol\beta_t^\intercal \mathbf{x}_{i,t}+
	\psi_j \eta_{i,t} , \sigma^2_j\right) $$
	for $i=1,\dots,n$. Here $y_{i,m_i}$ is the last observed gap time (in the $\log$ scale) for any $i$, and $y_{i, m_i+1}=\log(\tau_i-(\e^{y_{i,1}}+\cdots+\e^{y_{i,m_i}}))$ is the amount of time, in the $\log$ scale, between the censoring time and the time of the last observed event. 
	
	\item[{[3]}] Update the common regression coefficients: by conjugacy, the full-conditional is the multivariate $p_1$-dimensional Gaussian, with mean $\bm{\widetilde{\beta}}_0$ and variance-covariance matrix $\widetilde{\Sigma}_0$, 
	where 
	$$\widetilde{\Sigma}_0 = \left( \Sigma_0^{-1} + \sum_{j=1}^{ {k_n}} \frac{1}{\sigma_j^2}  \left( \sum_{i\in A_j}(m_i+1)\mathbf{x}_i \mathbf{x}_i^\intercal \right) \right)^{-1} $$ 
	and 
	$$ \widetilde{\boldsymbol\beta}_0 = \widetilde{\Sigma}_0\left(\sum_{i=1}^{ {k_n}} \sum_{i\in A_j}
	\sum_{t=1}^{m_i+1} \dfrac{y_{i,t} - (\alpha_j +  \boldsymbol\beta_t^\intercal \mathbf{x}_{i,t} + \psi_j \eta_{i,t})}{\sigma^2_j}\mathbf{x}_i \right).$$

	\item[{[4]}] Update the time-gap specific regression coefficients: each parameter vector $\bm\beta_t$ is sampled independently from the multivariate $p_2$-dimensional Gaussian, with mean $\widetilde{\bm\beta}_t$ and variance-covariance matrix $\widetilde{\Sigma}_t$, where
	$$\widetilde{\Sigma}_t = \left( \Xi_0^{-1} + \sum_{j=1}^{ {k_n}} \sum_{i\in A_j: m_i+1 \geq t} \dfrac{1 }{\sigma^2_j}\mathbf{x}_{i,t} \mathbf{x}_{i,t}^\intercal \right)^{-1} $$
	and
	$$ 	\widetilde{\bm\beta}_t = \widetilde{\Sigma}_t\left( \sum_{j=1}^{ {k_n}} \sum_{i\in A_j: m_i+1 \geq t}	\dfrac{y_{i,t} - (\alpha_l + \mathbf{x}_{i}^\intercal\boldsymbol\beta_0 + \psi_l \eta_{i,t})}{\sigma^2_l}\mathbf{x}_{i,t} \right),$$
 for $t = 1, \dots, J$. Here
 $\Xi_0$ is the diagonal matrix $\diag(\xi^2_1,\ldots,\xi^2_{p_2})$.

	\item[{[5]}] Update the dispersion parameter $\xi_1^2, \dots, \xi_{p_2}^2$: each parameter is independently sampled from 
	$$\xi^2_m\mid - \sim \mathrm{IG}\left(\nu_m , \gamma_m  \right),$$
	with $\nu_m = \nu_0 +\dfrac{J}{2} $ and $\gamma_m = \gamma_0 + \dfrac{1}{2}\sum_{t=1}^J\beta^2_{tm}$, for $m=1,\dots, p_2$.

	\item[{[6]}] Update the cluster specific parameters:	the likelihood for data in cluster $A_j$ that is used to build the joint distribution 
	for $(\alpha_j, \psi_j, \sigma^2_j)$ is proportional to
	$$\mathcal{L}( \bfy^*_j\mid -) \propto \prod_{i \in A_j} \prod_{t=1}^{m_i+1}  \dfrac{1}{\sqrt{2\pi \sigma^2_j}}\exp\left(-\dfrac{1}{2\sigma^2_j} \left(\hat{y}_{i,t} - (\alpha_j+\psi_j\eta_{i,t})\right)^2 \right) $$
	with $\hat{y}_{i,t} =  y_{i,t}- \boldsymbol\beta_t^\intercal \mathbf{x}_{i,t} -\boldsymbol\beta_0^\intercal \mathbf{x}_{i}$. It is straightforward to check that this is the likelihood of a regression model where the regression parameters are $\alpha_j,\psi_j$ and the residual variance is $\sigma_j^2$, with prior $P_0$ as specified in 
	\eqref{eq:prior_P0}, that is the usual conjugate prior.  
	Therefore, the full-conditionals these parameters are:
	\[
	\begin{split}
	\sigma^2_j\mid- &\sim \mathrm{IG}( \widetilde{a}_j, \widetilde{b}_j)\\
	(\alpha_j, \psi_j)^\intercal\mid\sigma_j^2,- &\sim \mathrm N_2(\widetilde{\bm{\theta}}_0, \sigma_j^2 \widetilde K_j)
	\end{split}
	\]
	where 
	\[
	\begin{split}
	\widetilde K_j &=\left( \sum_{i\in A_j}\sum_{t=1}^{m_i+1}\boldsymbol\zeta_{i,t}\boldsymbol\zeta_{i,t}^\intercal +  \diag(\kappa_0^{-1},\kappa_1^{-1})\right)^{-1}\\
	\widetilde{\boldsymbol\theta}_0 &= \widetilde K_j\left(  \sum_{i\in A_j}\sum_{t=1}^{m_i+1}\hat{y}_{i,t}\boldsymbol\zeta_{i,t} +  \diag(\kappa_0^{-1},\kappa_1^{-1}) \boldsymbol\theta_0  \right)\\
	\widetilde{a}_j &= a + \frac{n_j}{2}\\
	\widetilde{b}_j &= b + \dfrac{1}{2}\left( \sum_{i\in A_j}\sum_{t=1}^{m_i+1}\hat{y}_{i,t}^2 + \boldsymbol\theta_0^\intercal \diag(\kappa_0^{-1},\kappa_1^{-1})\boldsymbol\theta_0 - 
	\widetilde{\boldsymbol\theta}_0^\intercal \widetilde K_j^{-1}\widetilde{\boldsymbol\theta}_0\right)
	\end{split}
	\]
	and $m_j=\sum_{i\in A_j}(m_i+1)$, $\zeta_{i,t} = (1, \eta_{i,t})^\intercal$.

	\item[{[7]}] Update the latent partition of the data: we  adapt Algorithm 8 in \cite{Neal00} 
	to  consider  non-conjugacy of the kernel density and $P_0$   and to take into account the predictive structure of the PPMx-mixt prior.
 In the non-conjugate case,   the full conditionals of the cluster allocations  depend on a vector of  cluster-specific parameters. 
	The latter vector must be augmented by considering  $R$ new auxiliary variables $\left\{ \alpha_r, \psi_r, \sigma^2_r \right\}$ sampled from the prior $P_0$, representing  potential new clusters. To improve the mixing, this augmentation step is implemented adopting the re-use strategy in \cite{FavTeh13}.
	In particular, the probability of assigning the $i$-th subject to cluster $j$,  $j = 1,2,\dots,   k_n^{-i}$, similarly as in \eqref{eq:clusterassign_Gibbs}, here becomes
	\begin{equation}
	\label{eq:part_prob}
	\begin{split}
	\Pr\left( s_i=j\mid - \right) &\propto   \dfrac{c(u,n_j+1)g\left(\mathbf{x}^*_j \cup \mathbf{x}_i \right) }{c(u,n_j)g\left(\mathbf{x}^*_j \right)}
	\\
	& \qquad \times \prod_{t=1}^{m_i+1}\phi\left(y_{it}; \alpha_j + \mathbf{x}_{it}^\intercal \boldsymbol\beta_t +  \mathbf{x}_{i}^\intercal \boldsymbol\beta_0 + \psi_j\eta_{it},\sigma^2_j \right)
	\end{split}
	\end{equation}
	where $\rho_{n-1}^{-i}$ has the same definition as in the previous section. 
	{Analogously, observing that $g(\emptyset)=1$,  the probability of allocating the subject to 
	one of the new $R$ clusters is, for $ j=k_n^{-i}+1,...,k_n^{-i}+R$
	\begin{equation}
	\label{eq:part_prob_nre}
	\begin{split}
		\Pr\left( s_i=j\mid - \right) \propto \frac{1}{R}c(u,1)&g\left(\mathbf{x}_i \right) \prod_{t=1}^{m_i+1}\phi\left(y_{it}; \alpha_j + \mathbf{x}_{it}^\intercal \boldsymbol\beta_t +  \mathbf{x}_{i}^\intercal \boldsymbol\beta_0 + \psi_j\eta_{it},\sigma^2_j \right) 
	\end{split}
\end{equation}
	Specifically, under the NGG assumption, 
	$ \dfrac{c(u,n_j+1)}{c(u,n_j)} = \dfrac{n_j-\sigma}{(1+u)}$ if $n_j>0$, and $\kappa (u+1)^{\sigma}$ for $n_j=0$. 
}	
	
	\item[{[8]}] Update the latent parameter $u$: given the partition $\rho_n$, the auxiliary variable $u$ and data are independent, so that 
	$\mathcal{L}(u\mid-)$ $\propto$ $u^{n-1} \e^{-\Psi(u)} \prod_{j=1}^{k_n} c(u,n_j)$ with $\Psi(u)= \kappa \int_0^{+\infty} (1-\e^{-us}) \rho(s) ds$;  in the case 
	of the NGG process, this full-conditional simplifies to
	$$ \mathcal{L}(\D u\mid-) \propto \dfrac{u^{n-1}}{(u+1)^{n-\sigma k_n }}\e^{- \frac{\kappa}{\sigma}\left( \left( u+1\right)^{\sigma}-1\right)} \uno_{(0,+\infty)}(u)\D u.  $$
	In particular, to sample from this distribution, we
	use a simple Metropolis-Hastings update with a Gaussian proposal kernel truncated in $\left(0, +\infty\right)$.

\end{enumerate}

\section{Robusteness with respect to the similarity function}\label{sec:sim1}

 We provide a simulation study to compare the effect of the three similarity functions on posterior  distribution in a linear regression context. We have simulated  $n=200$ observations $(y_i,1, x_{i1},\ldots, x_{i4})$ for $i=1,\ldots,n$. 
The last two covariates are binary, while the first two are continuous. We generated data independently from three groups, with sizes 75, 75 and 50, respectively, as follows
\begin{equation*}
(x_{i1},x_{i2}) \iid \calN_2({\bm \mu_j},0.5\mathbb{I}_2), \quad
x_{i3},x_{i4} \iid \textsc{Bern}(q_j), \quad
Y_i \sim \calN({\mathbf x}_i^T{\bm\beta^0_j}, 0.5), \quad j=1,2,3. 
\end{equation*}
Specifically, 
\begin{enumerate}
	\item[1)] for the first group we set ${\bm \mu_1} =(-3,3)$, $q_1=0.1$, and 
	${\bm\beta^0_1} =(1,  5, 2, 1, 0)$;
	\item[2)] for the second group we have
	${\bm \mu_2} =(0,0)$, $q_2=0.5$, and ${\bm\beta^0_2} =(4,  2,-2, 1, -1)$;
	\item[3)] for the third group we have
	${\bm \mu_3} =(3,3)$, $q_3=0.9$, and ${\bm\beta^0_3} =(-1,-5,-2, -1, 1)$.
\end{enumerate}
\begin{figure}[!h]
	\centering
	
	\includegraphics[width = \textwidth]{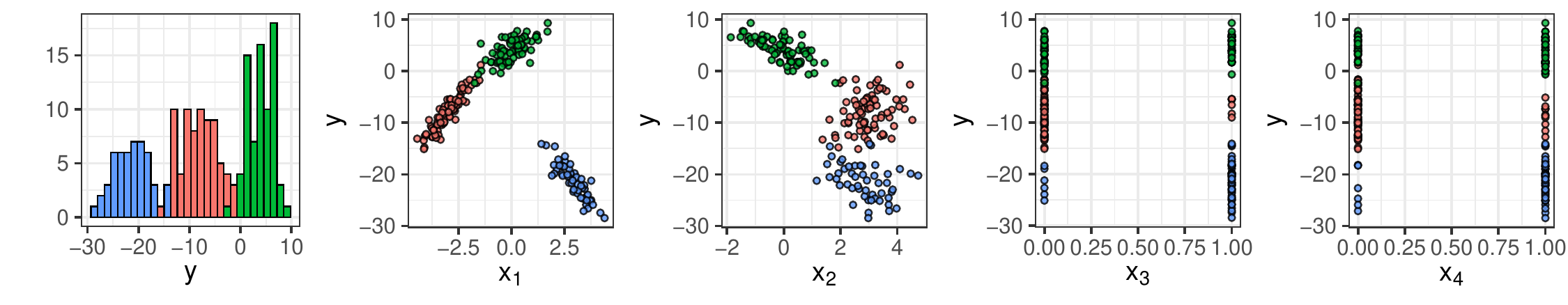}
	\caption{Simulated data. Left to right: histogram of the response variable $y_i$, scatterplots of   $x_{i1}$ and $x_{i2}$ (continuous covariates) and scatterplots of $x_{i3}$ and $x_{i4}$ (discrete covariates) versus the response variable. Different colours represent the three different groups   from which the data have been generated.}
	\label{fig:hist_simulated}
\end{figure}

Figure~\ref{fig:hist_simulated} shows the
simulated dataset. 
Three separate groups are clear for this figure, looking at either the responses and the covariates. We have fitted  model \eqref{eq:lik1}-\eqref{eq:rho_x}, with intensity given by a NGG process, when 
\begin{equation*}
f(\bfy^*_j\mid \bfx^*_j, \bm{\theta}_j^*) =\prod_{i\in A_j}
\phi(y_i; \mathbf{x}_i^\intercal \bm{\beta}_j,\sigma^2_j),
\end{equation*}
and $\phi(y_i; \mathbf{x}_i^\intercal \bm \beta_j,\sigma^2_j)$ is the univariate Gaussian density with mean $\mathbf{x}_i^\intercal\bm \beta_j$ and variance $\sigma^2_j$.
We include the whole vectors of $\mathbf{x}_i$ in the similarity and assume
the cohesion function of the NGG process with $\kappa=0.3$, $\sigma=0.2$, so that the prior number of clusters without covariate effect ($g\equiv 1$) has mean equal to 5.9 and variance equal to 7.7. 
Moreover, the base measure $P_0$ on $\mathbb R^p \times \mathbb R_{+}$ is $\mathrm{N}_p\left( \bm{0}, \sigma^2/\kappa_0 \mathbb{I}_{5\times 5}\right) \times\mathrm{IG}(a,b) $
with $\kappa_0 = 0.01$, $(a,b)=(2,1)$, and $\mathrm{IG}(a,b)$ denotes the inverse-gamma distribution with
mean $b/(a-1)$.

We run the algorithm described in Section~\ref{sec:ppmx_gibbs} of the Appendix to obtain $5\,000$ final iterations, after a burnin of $10\,000$.  computed the posterior estimated partition as the one $\rho_n$ minimizing the posterior expection of the variation of information loss function with equal missclassification costs \citep{Wad18,Ras18}. When classifying the datapoints according to this cluster estimates, we found that  missclassification rates  are $1\%$,   $2.5\%$ and  $8.5\%$ for 
$g_C$, $g_A$ ($\alpha=1$), and  $g\equiv 1$, respectively, when $\lambda=0.5$. 

\begin{figure}[!ht]
	\centering
	\includegraphics[width=0.32\textwidth,height=0.28\textwidth]{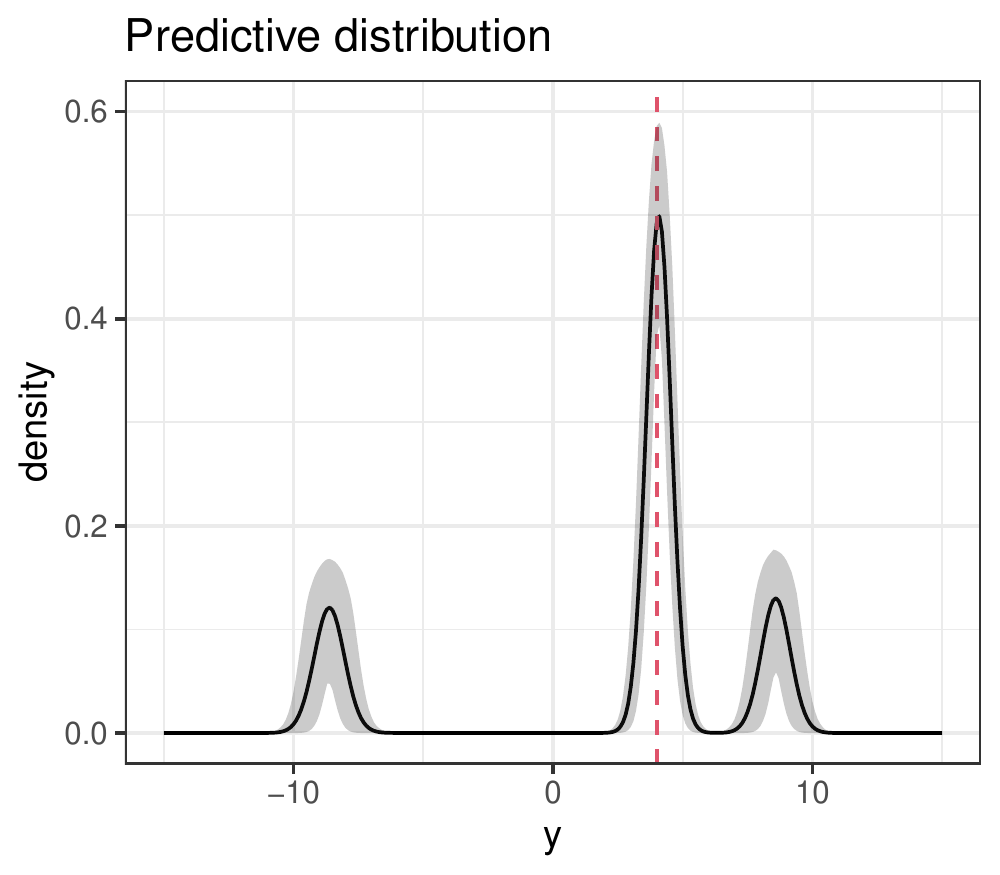}
	\includegraphics[width=0.32\textwidth,height=0.28\textwidth]{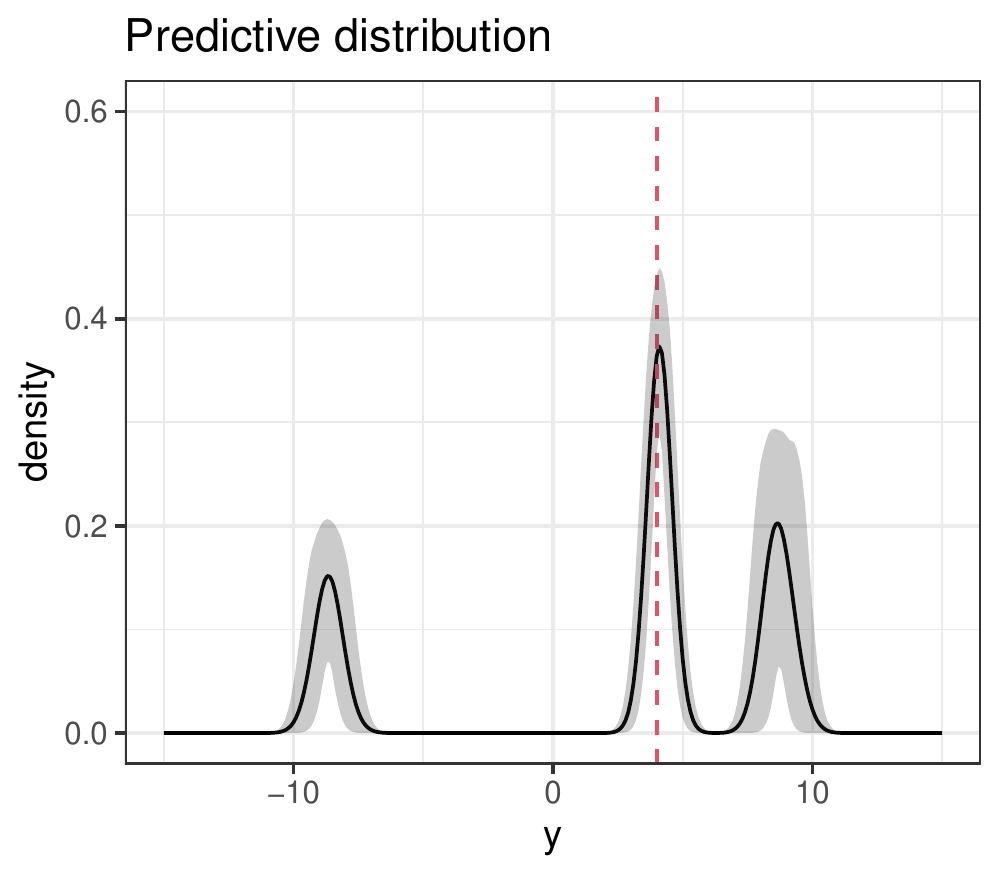}
	\includegraphics[width=0.32\textwidth,height=0.28\textwidth]{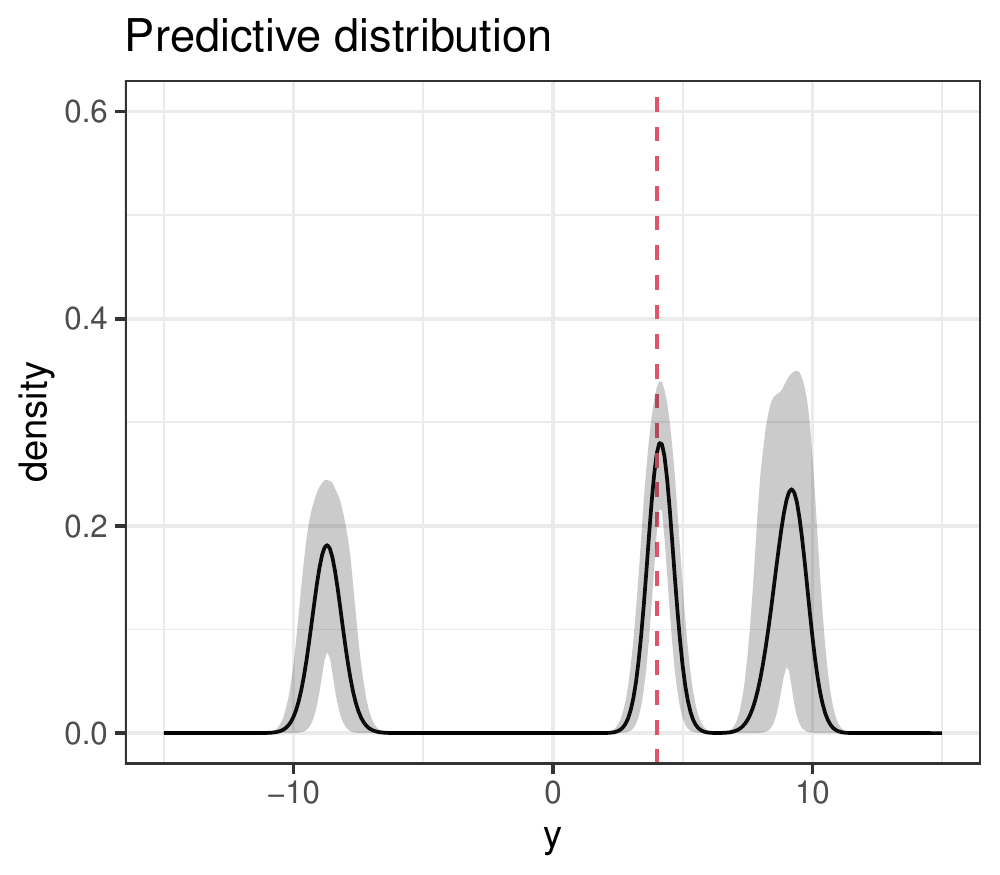}
	\caption{Posterior predictive distribution of $Y_1$,   i.e., for a \textit{new} individual with the same covariate values as the first individual in the sample,   under $g_C$ (left), $g_A$ (center) and $g\equiv 1$ (right). The dashed red vertical lines denote the true value.   Black lines denote the posterior mean of the density, the shaded areas denote 90\% credibility band,   based on the MCMC sample.}
	\label{fig:pred_simulated}
\end{figure}
\begin{figure}[!h]
	\centering
	\includegraphics[width = \textwidth]{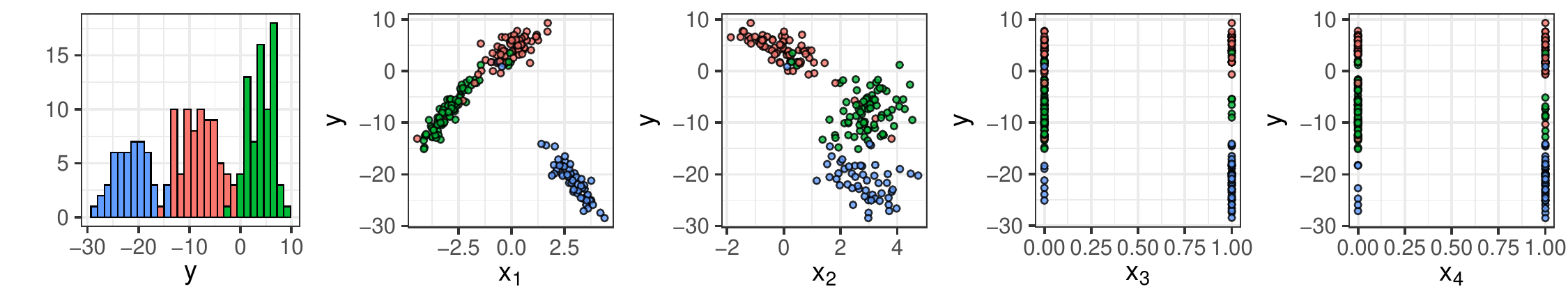}
	\caption{Simulated data. Left to right: histogram of the response variable $y_i$, scatterplots of   $x_{i1}$ and $x_{i2}$ (continuous covariates) and scatterplots of $x_{i3}$ and $x_{i4}$ (discrete covariates) versus the response variable.  Different colours represent the different clusters estimated with the   PPMx-mixt model and $g_C$ is the similarity function in the prior.}
	\label{fig:clusterestimategA_simulated}
\end{figure}
\begin{figure}[!h]
	\centering
	\includegraphics[width = \textwidth]{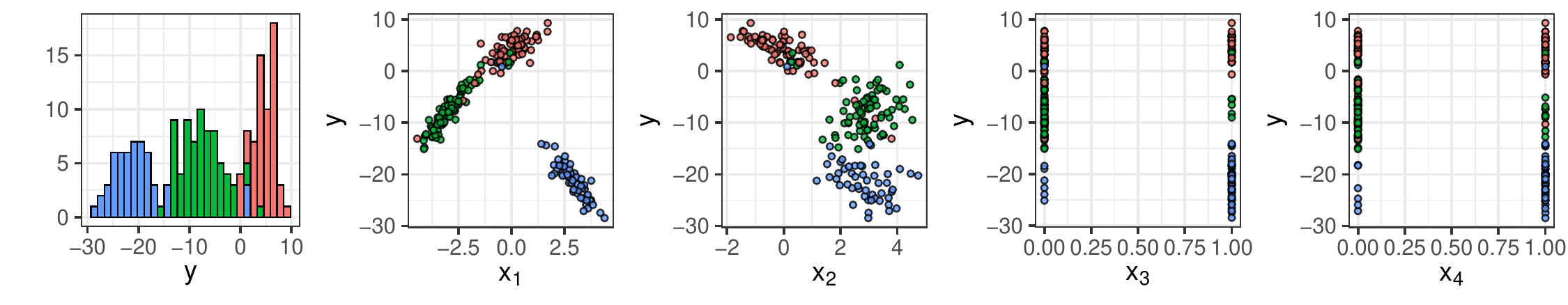}
	\caption{Simulated data. Left to right: histogram of the response variable $y_i$, scatterplots of   $x_{i1}$ and $x_{i2}$ (continuous covariates) and scatterplots of $x_{i3}$ and $x_{i4}$ (discrete covariates) versus the response variable. Different colours represent the different clusters estimated with the   PPM prior without covariate in the prior ($g\equiv 1$).  }
	\label{fig:clusterestimateNOCOV_simulated}
\end{figure}

We have also computed posterior predictive distributions as explained at the end of Section~\ref{sec:posterior_calc}. 
Figure~\ref{fig:pred_simulated} reports the posterior predictive distribution for a \textit{new} individual, 
with the same covariate vector as the first subject in the sample. It is clear from the figure that the prediction is 
more precise under $g_A$ and $g_C$. In fact, when  
we do not include covariate information in the prior for the random partition, 
the posterior predictive density is not able to distinguish to which of the three groups the item belongs and the three peaks have approximately the same height. 
In contrast, when $g_A$ or $g_C$ are included in the prior, 
the posterior predictive density exhibits one main peak, so that covariate information helps, in this case, in selecting the right group the observation should be assigned to.
This is also confirmed by the missclassification rates we have reported above.
 Figures~\ref{fig:clusterestimategA_simulated} and \ref{fig:clusterestimateNOCOV_simulated} report the  cluster estimate under $g_C$ and $g\equiv 1$ (no covariates in the prior), respectively. It is  clear also from these plots  that the inclusion of covariates in the prior specification improves the clustering performance of the model. 




\section{Comparison to alternative models}\label{sec:sim2}

 We fit model \eqref{eq:lik1}-\eqref{eq:rho_x}, with intensity given by a NGG process, in the regression context,   to the same simulated dataset as in \cite{PPMx_JCGS11}, Section 5.2.
The simulation \virgolette{truth} consists
of 12 different distributions, corresponding to different
covariate settings (see Figure 1 of that paper).  The original dataset contains 1,000 data with three covariates $\mathbf{x}_i:=(x_{i1}, x_{i2}, x_{i3})$ for each item $i$, where $x_{i1}\in\{-1,0,1\}$ and $x_{i2}$ and  $x_{i3}$ are binary. We adopt the conditional distribution of data in cluster $A_j$ (see \eqref{eq:lik1}) as 
\begin{equation*}
f(\bfy_j^*\mid \bfx^*_j, \bm{\theta}_j^*) =\prod_{i\in A_j}
\phi(y_i; \mathbf{x}_i^\intercal \bm \beta_j,\sigma^2_j),
\end{equation*}
where $\phi(y_i; \mathbf{x}_i^\intercal \bm \beta_j,\sigma^2_j)$ is the univariate Gaussian density with mean $\mathbf{x}_i^\intercal\bm \beta_j$ and variance $\sigma^2_j$. The linear term includes the intercept. We assume the prior as in Section~\ref{sec:model}, with similarity function $g_C$   and distance $d(\mathbf{x}_1,\mathbf{x}_2)$ as described in Section~\ref{sec:similarity_g}.
The prior $P_0$ of the cluster-specific parameters is a normal-inverse-gamma distribution, with
\begin{equation*}
\bm (\beta_j,\sigma^2_j) \sim \calN_4(\bm{\beta}_0,\sigma^2_j/\kappa_0 \mathbb{I}_{4\times 4} ) \times \mathrm{IG}(\sigma^2_j; a_0,b_0).  
\end{equation*}
To replicate tests in \cite{PPMx_JCGS11} and \cite{DPP} a total of 
$M = 100$ datasets of size 200 were generated
by randomly subsampling 200 out of the 1,000 available
observations. We compute the root MSE (over the 100 datasets) 
for estimating $\E(Y\mid  x_1; x_2; x_3)$ for each of the 12 covariate
combinations \citep[see details in][]{PPMx_JCGS11}. 
We fix $P_0$ according via the empirical Bayes approach using overall sample mean and variance of the responses. 
Table \ref{tab:mse_12cov} displays root MSE under our model when the hyperparameters in the cohesion function are fixed as $\sigma=0.2$, $\kappa= 0.001$; this corresponds assuming that, without covariate effect (i.e. when $g\equiv 1$),  the prior number of clusters has mean equal to 3 and variance equal to 5.8. 
Parameter $\lambda$ in $g_C$ has been fixed to 0.041, with $\varepsilon^*$ in Section \ref{sec:similarity_g} equal to 0.1; the table reports also  the last two columns in Table 7 in \cite{DPP}. 

\begin{table}[h!]
	\begin{center}\small
		\begin{tabular}{c c c c c c }
			\hline
			$x_1$ & $x_2$ & $x_3$ & PPMx-mixt   & DPP & PPMx \\ \hline
			-1 & 0 & 0 & 14.0 & 6.1 & 7.9 \\
			0 & 0 & 0 & 2.8 & 6.7 & 3.9 \\
			1 & 0 & 0 & 3.1 & 7.2 & 2.8 \\
			-1 & 1 & 0 & 13.9 & 6.5 & 5.4 \\
			0 & 1 & 0 & 3.3 & 6.5 & 4.6 \\
			1 & 1 & 0 & 8.7 & 6.8 & 4.0 \\
			-1 & 0 & 1 & 5.9 & 6.8 & 6.1 \\
			0 & 0 & 1 & 2.2 & 6.1 & 4.2 \\
			1 & 0 & 1 & 2.2 & 5.7 & 4.5 \\
			-1 & 1 & 1 & 4.8 & 5.9 & 9.5 \\
			0 & 1 & 1 & 2.6 & 6.6 & 8.3 \\
			1 & 1 & 1 & 2.4 & 5.8 & 6.2 \\ \hline
			\textbf{avg} & & & \textbf{5.5} &\textbf{6.4} & \textbf{5.6} \\
			\hline
		\end{tabular}
	\end{center}
	\caption{Root MSE for estimating $\mathbb{E}(Y\mid x_1 , x_2 , x_3 )$ for
		12 combinations of covariates $\left(x_1 , x_2 , x_3 \right)$ and $PPM_x$ and DPP 
		as competing models of reference; the last two columns are those in Table~7 of
		\citealp{DPP}.
	}
	\label{tab:mse_12cov}
\end{table}

The PPMx-mixt shows performance comparable to the competitors DPP and PPMx in \cite{DPP} and \cite{PPMx_JCGS11}, respectively. Note that our model gives more variability among different combinations of covariates.

\end{document}